\input{psfig.sty}
\documentclass[useAMS]{mn2e}
\usepackage{epsfig}
\usepackage{graphicx}
\usepackage{color}
\usepackage{subfigure}
\newcommand{\goodgap}{\hspace{\subfigtopskip} \hspace{\subfigbottomskip}}

\title[Density profile slope in Dwarfs and environment]{Density profile slope in Dwarfs and environment}
\author[A. Del Popolo]{A. Del Popolo$^{1, 2}$\thanks{E-mail:
antonino.delpopolo@unibg.it} \\
$^{1}$Dipartimento di Fisica e Astronomia, Universit\'a di Catania, Viale Andrea Doria 6, 95125 Catania, Italy\\
$^{2}$Argelander-Institut f\"ur Astronomie, Auf dem H\"ugel 71, D-53121 Bonn, Germany}
\begin{document}

%\date{Accepted 1988 December 15. Received 1988 December 14; in original form 1988 October 11}

\pagerange{\pageref{firstpage}--\pageref{lastpage}} \pubyear{2002}

\maketitle

\label{firstpage}

\begin{abstract}
In the present paper, we study how the dark matter density profiles of dwarfs galaxies in the mass range $10^8-10^{10} M_{\odot}$ are modified 
by the interaction of the dwarf in study with the neighboring structures, 
and by changing baryon fraction in 
dwarfs.
To this aim, we determine the density profiles of the quoted dwarfs by means of Del Popolo (2009) which takes into account the effect of tidal interaction with 
neighboring structures, those of ordered and random angular momentum, dynamical friction, response of dark matter halos to condensation of baryons, and effects 
produced by baryons presence. As already shown in Del Popolo (2009), the slope of density profile of inner halos flattens with decreasing halo mass and the profile is well approximated by a Burkert profile. We then treat angular momentum generated by tidal torques and baryon fraction as a parameter in order to understand 
how the last influences the density profiles. 
The analysis shows that 
dwarfs who suffered a smaller tidal torquing (consequently having smaller angular momentum) are characterized by steeper profiles with respect to 
dwarfs subject to higher torque, and similarly dwarfs having a smaller baryons fraction have also steeper profiles than those having a larger baryon fraction.
In the case tidal torquing is shut down 
and baryons are not present, the density profile is very well approximated by an Einasto profile, similarly to dwarfs obtained in dissipationless N-body simulations. 
We then apply the result of the previous analysis to the dark matter halo rotation curves of three different dwarfs, namely NGC 2976, known to have a flat inner core, NGC 5949 having a profile intermediate between a cored and a cuspy one, and  NGC 5963 having a cuspy profile. After calculating baryon fraction, which is $\simeq 0.1$ for the three galaxies, we fitted the rotation curves 
changing the value of angular momentum. NGC 2976, has an higher value of ordered angular momentum ($\lambda \simeq 0.04$) with respect to NGC 5949 ($\lambda \simeq 0.025$) and in the case of NGC 5963 the very steep profile can be obtained with a low value of $\lambda$ ($\lambda \simeq 0.02$) and also decreasing the value of the random angular momentum. In the case of NGC 2976 tidal interaction with M81 could have also influenced the inner part of the density profile. We finally show how the inner density profile correlates with Karachentsev et al. (2004), and Karachentsev and Kashibadze (2006) tidal index for dwarfs and LSBs.

\end{abstract}

\begin{keywords}
cosmology--theory--large scale structure of Universe--galaxies--formation
%circumstellar matter -- infrared: stars.
\end{keywords}

\section{Introduction}

Notwithstanding dwarf galaxies are the most diffused type of galaxies in the Universe, they have received less attention than spiral and elliptical galaxies. Their relevance is noteworthy in cosmology since their study may bring new insights in the processes present in the early universe and since the dwarfs that we observe today are probably the survivors of the 
%constituents 
building blocks of larger galaxies. Dwarfs can be classified in dwarf spiral galaxies, blue compact dwarf galaxies  (BCDs), dwarf irregular galaxies (dIrrs), dwarf elliptical galaxies (dEs), dwarf spheroidal galaxies (dSphs), and tidal dwarf galaxies. They are all characterized by small size, low mass, low metallicity, and low luminosity, and are moreover characterized by an old population (Grebel 2001). The so called morphological segregation 
%or as the morphology--density relation
is the difference observed in spatial distribution of dSphs, DEs, and dIrrs, and is probably originated by the effects that environment has in the evolution of galaxies.
As previous reported, dwarfs are important cosmological probes and it is not an incident that the missing satellites problem (Klypin et al. 1999; Moore et al. 1999) and the cusp-core problem, (de Blok 2010), namely the discrepancy of observed and simulated density profiles in the inner kpc of the halo, are to them connected. For what concerns the second problem, on which we shall concentrate in the rest of the paper, comparison of rotation theoretical curves, obtained by Moore (1994) and Flores \& Primack (1994), with HI rotation curves of late-type, gas-rich, dwarf galaxies lead to the conclusion that these objects are dark matter dominated and their halos are best fitted by constant-density core, instead of the theoretical predicted cuspy profiles. This lead to the conclusion that CDM predictions on small scales are incompatible with observations (Carignan \& Beaulieu 1989;  
%Lake, Schommer, and van Gorkom 1990; Jobin \& Carignan 1990; 
Broeils 1990) of dwarf galaxies (Moore 1994). Also LSB galaxies seem to indicate that the shape of the density profile at small scales is significantly shallower than what is found in numerical simulations (e.g., de Blok, Bosma, and McGaugh 2003), and a similar indication is given by spheroidal galaxies (e.g., Ursa Minor (Kleyna et al. 2003), Draco dSph (Magorrian 2003)). 

According to the studies of the density profile of dwarfs, the inner part of density profiles is characterized by a core-like structure (e.g., Flores \& Primak 1994; Moore 1994;  Kravtsov et al. 1998;  de Blok \& Bosma 2002 (hereafter dBB02); de Blok, Bosma \& McGaugh 2003; Gentile et al. 2004, 2006; Blaise-Ouelette et al. 2004,  Span\'o et al. 2008, Kuzio de Naray et al. 2008, 2009, and Oh et al. 2010 (hereafter Oh10)),
%(Flores \& Primak 1994; Moore 1994; Burkert 1995; Kravtsov et al. 1998; Salucci \& Burkert 2000; Borriello \& Salucci 2001; de Blok \& Bosma 2002; Marchesini et al. 2002; %de Blok 2003; de Blok, Bosma \& McGaugh 2003; Gentile et al. 2004, 2006; Blaise-Ouelette et al. 2004,  Span\'o et al. 2008, Kuzio de Naray et al. 2008, 2009, and Oh et al. %2010), 
but some studies 
found that density profiles are cuspy. For example, Hayashi et al. (2004) found that about 70\% of galaxies in the sample they used have rotation curves 
consistent with steep slopes. Other studies concluded that density profiles are compatible with both cuspy and cored profiles (van den Bosch et al. 2000; Swaters et al. 2003; Spekkens et al. 2005; Simon et al. 2005 (hereafter S05); de Blok et al. 2008 (hereafter dB08)). In the case of NGC 2976, 4605, 5949, 5693, 6689, S05 found that the power-law index, $\alpha$, describing the central density profile, span the range from $\alpha = 0$ (NGC 2976) to $\alpha = 1.28$ (NGC 5963), with a 
mean value $\alpha= 0.73$, shallower than that predicted by the simulations, and scatter in $\alpha$ from galaxy to galaxy is equal to 0.44, which is 3 times as large as in cold dark matter (CDM) simulations. dB08, by using the THINGS sample, found that for galaxies having $M_B < -19$ a NFW profile or a pseudo-isothermal (PI) profile statistically fit equally well the density profiles of galaxies, while for $M_B > -19$ the core dominated PI model fits significantly better than the NFW model. In other terms, for low mass galaxies a core dominated halo is clearly preferred over a cusp-like halo, while for massive, disk dominated galaxies, the two models fit apparently equally well.
%The previous discussion pointed out that, although large part of the observations conclude that dwarfs have a core-like profile (e.g., de Blok, Bosma \& McGaugh 2003), this %does not imply that all dwarfs have core-like profiles (e.g., S05).

The previous arguments lead to some important questions: is there a correlation between the value of slope of galaxies tidally influenced by external galaxies and the tidal field? How much the baryons in dwarfs influence their final density profiles? Since some studies (e.g., S05, dB08) hint to a possible correlation between inner slope of dwarfs and their mass and interaction with neighbors, would be important to verify from a theoretical point of view if the quoted correlations are admissible. On the other hand, the morphology--density relation illustrates that environment is an important factor in dwarfs evolution.

%In order to verify if the quoted correlation is real, we have to improve 

This paper has a double goal: the first one is to see how changes in baryons fraction of galaxies and magnitude of tidal interaction modify the density profile of galaxy. It is important to stress that in Del Popolo (2009) the effects of baryon presence was already taken into account, but we used fixed values of baryon fraction, according to the type of galaxies studied, and angular momentum was obtained and fixed to the value given by the  tidal torque theory (TTT). While nowadays we are able to compute with reasonable precision the evolution of dark matter component, physical processes driving galaxy formation, like the evolution of baryonic matter, cooling, star formation, black hole formation, and SN and AGN feedback, are extremely complex and still debated. However, these processes, usually referred to as sub-grid physics, can be taken into account with new computer models. One example is the hydrodynamical simulation presented by Governato et al. (2010) (hereafter G10), which obtained bulgeless dwarf galaxies with shallow inner dark mater profile. In their study, the authors took account of baryons and showed how supernovae explosions affect galaxy evolution. The result obtained needed a intensive use of supercomputers for long CPU times. 
%The reason is due to the fact that the point of force of numerical simulations (namely,
Analytical and semi-analytical models, like the one presented in Del Popolo (2009), have several advantages when compared to simulations: a) flexibility (one can study the effects of physical processes one at a time including subgrid processes); b) computational efficiency (it takes about 10 s to compute the density profile of a given object at a given epoch on a desktop PC, Ascasibar et al. 2007). If we compare our model to G10 simulation, as we will show in the following, we obtain 
accurate density profiles,  but at the same time we can obtain different density profiles for different values of angular momentum and baryon fraction, with small CPU times. Studying how baryon fraction or tidal interaction with large neighbors influence the density profile, by means of G10 simulations, would imply to perform each time a new simulation, with a prohibitive computational cost\footnote{For this reason, in Governato et al. (2010) simulation were fundamentally calculated two isolated haloes with a fixed value of baryon fraction.}.

The second goal of the present paper is that of using the results of the first goal (changes of density profiles with baryon fraction and angular momentum) 
in order to clarify if the density profile of some observed galaxies with cuspy, cored, and intermediate density profiles can be re-obtained in our model.
%The first goal of the present paper and its results will be used to implement the second goal, namely to see if the density profile of some observed galaxies with cuspy, %cored, and intermediate density profiles can be re-obtained in our model.
So, in the present paper, we analyze how different tidal torques and baryons content of haloes influence dwarfs 
density profile. 

The paper is organized as follows: in Section 2, we summarize the model used. In Section 3.1, we determine the dwarfs density profiles when tidal torques and baryons fraction is changed. In Section 3.2, we compare the previous results with three different galaxies,
namely NGC 2976, NGC 5949, and NGC 5963, we discuss why the three dwarfs have different profiles, and finally extend the study to some other galaxies.
Section 4 is devoted to discussions and conclusions.  

\section{Summary of Del Popolo 2009 model}

%\subsection{}

In the present study, the dwarfs galaxies dark matter haloes are formed using the analytical method introduced in Del Popolo (2009), which is summarized in the following.
We added a discussion on the way baryon fraction was calculated, not present in Del Popolo (2009) (hereafter DP09).

The spherical infall model (SIM), was studied for the first time by Gunn \& Gott (1972) and widely discussed and improved in a large series of following papers (Fillmore \& Goldreich 1984 (hereafter FG84);  Bertschinger 1985; Hoffman \& Shaham 1985 (hereafter HS); Ryden \& Gunn 1987 (hereafter RG87); White \&Zaritsky 1992; Avila-Reese et al. 1998; Nusser \& Sheth 1999; Henriksen \& Widrow 1999; Subramanian et al. 2000; Del Popolo et al. 2000; Hiotelis 2002; Le Delliou \& Henriksen 2003; Ascasibar et al. 2004; Williams et al. 2004).
In SIM one considers a spherical overdensity in the expanding universe, that behaves exactly as a closed sub-universe because of Birkhoff's theorem, and studies its evolution. 
A bound mass shell of the quoted overdensity will expand from an initial radius 
$x_i$ to a maximum one, $x_m$, which is usually termed apapsis or radius of turn-around, $x_{ta}$ (Peebles 1980):
\begin{equation}
x_m=g(x_i)=x_i\frac{1+{\overline \delta_i}}{{\overline \delta_i}-(\Omega_i^{-1}-1)}
\label{xtr}
\end{equation}
where $\Omega_i$ is the density parameter at time $t_i$, and the mean fractional density excess inside the shell, as measured at current epoch $t_0$, assuming linear growth, can be calculated as:
\begin{equation}
{\overline \delta_i}=\frac{3}{x_i^3} \int_0^{x_{i}} \delta(y)y^2 dy
\label{eq:overd}
\end{equation}
Eq. (\ref{eq:overd}) furnish a relation among the final time averaged radius of a given Lagrangian shell and its initial radius. 

The initial overdensity $\overline \delta_i (x_i)$ can be calculated once a spectrum of perturbations is known.
Throughout Del Popolo (2009), we adopted a $\Lambda$CDM cosmology with WMAP3 parameters, $\Omega_m=1-\Omega_{\Lambda}=0.24$,  $\Omega_{\Lambda}=0.76$, $\Omega_b=0.043$ and $h=0.73$, where $h$ is the Hubble constant in units of 100 km $s^{-1}$ $Mpc^{-1}$. The variance $\sigma_8=0.76$ (Romano-Diaz et al. 2008)
of the density field convolved with the top hat window of radius 8 $h^{-1}$ $Mpc^{-1}$ was used to normalize the power spectrum. 

Expressing the scaling of the final radius, $x$, with the initial one by relating $x$ to the turn around radius, $x_m$, the final radius can be expressed as:
\begin{equation}
x = f(x_i)x_m
\end{equation}
being $f$ the collapse parameter. Using mass conservation, and assuming that each shell is kept at its turn-around radius, the shape of the density profile is given by (Peebles 1980; HS; White \& Zaritsky 1992):
\begin{equation}
\rho_{ta}(x_m)=\rho_i (x_i) \left( \frac{x_i}{x_m} \right)^2 \frac{d x_i}{dx_m}
\label{eq:dturn}
\end{equation}

After reaching maximum radius, a shell collapses and will start oscillating and it will contribute to the inner shells with the result that energy is not  an integral of motion anymore and the collapse factor, $f$, is no longer constant. The following evolution of the system can be described 
assuming that the potential well near the center varies adiabatically (Gunn 1977, FG84).
If a shell has an apapsis radius (i.e., apocenter) $x_m$ and initial radius $x_i$, then the mass inside $x_m$ is obtained summing the mass contained in shells with apapsis smaller than $x_m$ (permanent component, $m_p$) and the contribution of the outer shells passing
momentarily through the shell $x_m$ (additional mass $m_{add}$), and can be expressed as: 
%The mass inside $x_m$, using mass conservation, can be expressed as:
\begin{eqnarray}
m_T(x_m)&=&m_p(x_m)+m_{add}(x_m)= \frac{4}{3} \pi \rho_{b,i} x_i^3 (1+{\overline \delta_i})\\ & &
+\int_{x_m}^{R} P_{r_m} (x) \frac{d m(x)}{dx} dx
\label{eq:mpp}
\end{eqnarray}
where $\rho_{b,i}$ is the initial background density,  $R$ is the radius of the system (the apapsis of the outer shell) 
and the distribution of mass $m(x)=m(x_m)$ is given by Eq. (\ref{eq:dturn}), while $P_{x_m}(x)$ is the probability to find the shell with apapsis $x$ inside radius $x_m$,
calculated as the ratio of the time the outer shell (with apapsis $x$) spends inside radius $x_m$ to its period.  

This last quantity can be expressed as 
\begin{equation}
P_{x_m}(x)= \frac{
\int_{x_p}^{x_m} \frac{d \eta}{v_x(\eta)} }
{\int_{x_p}^{x} \frac{d \eta}{v_x(\eta)}
}
\end{equation}
where $x_p$ is the pericenter of the shell with apsis $x$ and $v_x (\eta)$ is the radial velocity of the shell with apapsis $x$ as it passes from radius $\eta$.

This radial velocity can be obtained by integrating the equation of motion of the shell:
\begin{equation}
\frac{dv_r}{dt}=\frac{h^2(r,\nu )+j^2(r, \nu)}{r^3}-G(r) -\mu \frac{dr}{dt}+ \frac{\Lambda}{3}r 
\label{eq:coll}
\end{equation}
where $h(r,\nu )$ \footnote{
%As defined in subsection 3.1, 
$\nu=\delta(0)/\sigma$, where $\sigma$ is the mass variance filtered on a scale $R_f$.}
is the ordered specific angular momentum generated by tidal torques, $j(r, \nu)$ the random angular momentum (see RG87 and Del Popolo 2009), $G(r)$ the gravitational acceleration, $\Lambda$ the cosmological constant and $\mu$ the coefficient of dynamical friction.
The final density profile  is given by (Eq. A18, DP 2009):
\begin{equation}
\rho(x)=\frac{\rho_{ta}(x_m)}{f^3} \left[1+\frac{d \ln f}{d \ln x_m} \right]^{-1}
\label{eq:dturnnn}
\end{equation}

%CONTINUARE E GUARDARE QUELLO DI PRIMA \\

As seen by the previous discussion, the previous model (i.e., Del Popolo 2009) takes into account several effects, such as ordered
angular momentum, $h$, and random angular momentum, $j$, dynamical friction 
%$L_{rand}$\footnote{In the following, we use the specific ordered and random angular momentum, which shall be indicated with $h$, and $j$, respectively}, dynamical friction, 
and adiabatic contraction of dark matter (AC).  

Ordered angular momentum is due to the tidal torques, $\tau$, exerted by a field of random density fluctuations on proto-halos (Hoyle 1953; Peebles 1969; White 1984; Ryden 1988; Eisenstein \& Loeb 1995; Catelan \& Theuns 1996), it can be calculated when the r.m.s. torque, $\tau(r)$, is known, by integrating torque over time (see Del Popolo 2009 Appendix C, for details). 
Before going on, we recall that common practice is that of expressing the total angular momentum 
in terms of the dimensionless spin parameter
\begin{equation}
\lambda=\frac{L |E|^{1/2}}{GM^{5/2}}, 
\end{equation}
where $L$ is the angular momentum, summed over shells, and $E$ is the binding energy of the halo. As several authors showed (e.g., Vivitska et al. 2002 ), the $\lambda$ parameter is well described by a log-normal distribution (e.g. Vivitska et a. 2002), with a maximum of $\lambda=0.035$ and  90\% probability that $\lambda$ is in the range 0.02-0.1 (Vivitska et al. 2002). While Catelan \& Theuns 1996 showed a dependency of $\lambda$ on the peak height\footnote{$\nu$ is the central peak height given by $\nu=\overline{\delta}(0)/\sigma_0(R)$, where $\overline{\delta}(r)$ is the overdensity in the radius $r$, and $\sigma_0(R)$ the mass variance filtered on scale $R$. }
Knebe \& Power (2008) and Cervantes-Sodi at al. (2008a,b) (see Fig. 5 of Cervantes-Sodi et al. 2008a; and Fig. 3 of Cervantes-Sodi et al. 2008) showed a dependence on environment, cosmology, and mass, and Cervantes-Sodi et al. (2010b) found a correlation between $\lambda$ and cluster environment. 
So, while the baryons in bright disc galaxies seem to have spin parameters similar to those of their host haloes, dwarfs and LSB disc galaxies may tend to be associated with a higher spin parameter. For example, van den Bosch, Burkert \& Swaters (2001, hereafter BBS) studied the spin in a sample of 14 low-mass disc galaxies 
with an estimated average of virial velocity $V \simeq 60$ km/s. They found an average spin parameter about 50\% larger than that of the dark haloes (see also
Maller \& Dekel 2002, Fig. 8 and Dekel \& Woo 2003). 
Also Boissier et al. (2003), showed in their study that 35\% of all galaxies (in number) with $0.06<\lambda<0.21$ are LSBs (see also Jimenez et al. 1998). 
According to several analytical papers (Nusser 2001; Hiotelis 2002; Le Delliou \& Henriksen 2003; Ascasibar et al. 2004; Williams et al. 2004), objects with larger angular momentum should have flatter profiles, one therefore expects that these high spin objects should typically have shallower density cusps.
Moreover the central parts of these galaxies are also much less dense than the simulations indicate. 

Random angular momentum (RG87, see Appendix C2 of Del Popolo 2009), is connected to random velocities (see RG87), and in all previously quoted papers only this kind of angular momentum was taken into consideration. The usual approach (Nusser 2001; Hiotelis 2002; Ascasibar et al. 2003) consists in assigning a specific angular momentum at turn around:
\begin{equation}
j \propto \sqrt{GM r_m}  
\end{equation}
With this prescription, the orbital eccentricity $e$ is the same for all particles in the halo (Nusser 2001).
We assigned random angular momentum to protostructures expressing the specific angular momentum $j$ through the ratio 
%$e_0=\left( \frac{r_{min}}{r_{max}} \right)_0$, 
$e=\left( \frac{r_{min}}{r_{max}} \right)$, 
and left this quantity as a free parameter (Avila-Reese et al. 1998). This procedure is justified by N-body simulations of halo collapse:
for CDM halos, N-body simulations produce constant $< \frac{r_{min}}{r_{max}}> \simeq 0.2$ ratios of dark matter particles in virialized haloes.
%\footnote{The constancy of $< \frac{r_{min}}{r_{max}}>$ throughout the halo has been interpreted as a proof that the adiabatic approximation is largely valid for haloes %generated in N-body simulations (Jesseit et al. 2002).}, and the range for this value is around 0.1-0.3 (e.g., Ghigna et al. 1999).
Ascasibar et al. (2007), showed that there is a dependence on the quoted ratio on the dynamical state of the system: 
%dynamical state: major mergers are well described by constant eccentricity up to the virial radius \footnote{For their dark matter haloes, $r_{v}/r_{ta}$ is typically of %the order of 0.2-0.3.},  
major mergers are well described by constant eccentricity up to the virial radius, 
%\footnote{For their dark matter haloes, $r_{v}/r_{ta}$ is typically of the order of 0.2-0.3.}, 
relaxed systems are more consistent with a power-law 
profile. Minor mergers are in the middle. 
%The average profile can be fitted by a power law, but the slope is shallower than for relaxed systems. A least-square fit to the relaxed population yields:
%\begin{equation}
%e(r_{max}) \simeq 0.8 (r_{max}/r_{ta})^{0.1}
%\end{equation}
%for $r_{max}< 0.1 r_{ta}$. 
In the present paper, we use Avila-Reese et al. (1998) method with the correction of Ascasibar et al. (2003) (see their section 3.2 and their Eq. 32).

Dynamical friction was taken into account by introducing the dynamical friction force
(see Appendix D of Del Popolo 2009) in the equation of motions previously described.
%in Antonuccio-Delogu \& Colafrancesco (1994) (see also Appendix D of DP09). 

We also took into account adiabatic contraction (AC) of dark matter halos in response to the condensation of baryons in their centers, leading to a steepening of the dark matter density slope. Blumenthal et al. (1986) described an approximate analytical  model to calculate the effects of AC. More recently Gnedin et al. (2004) proposed a simple modification of the Blumenthal model, which describes numerical results more accurately. 
For systems in which angular momentum is exchanged between baryons and dark matter (e.g., through dynamical friction),
Klypin et al. (2002) introduced a modification to Blumenthal's model. 
The adiabatic contraction was taken into account by means of Gnedin et al. (2004) model, and Klypin et al. (2002) model was used to take also account of exchange of angular momentum between baryons and dark matter. 

For what concerns how baryon fraction and baryon distribution was chosen, we want to recall that the model of Del Popolo (2009) contained both DM and baryons. In that paper and in its Appendix E, we discussed the quantity of baryons introduced in the model and how their initial mass distribution and the final distribution was fixed. 
%In the present paper, 
Here, we speak again of baryons and baryon fraction, 
%and in a subsection distinct from the previous, 
since here we study how the density profiles changes in terms of change of the baryon fraction. Till now, almost all N-body simulations on density profiles have dealt just with dark matter halos. However, we know that 
halos contain baryons, that influence the evolution of dark matter. The effect of baryons on dark matter haloes was studied in El-Zant et al. (2001) and Romano-Diaz et al. (2008) simulation. 
Full SPH simulations have been performed recently by G10, showing the strong effect of baryon physics on the 
final dark matter profile. 
%For what concerns, semi-analytical models, Blumenthal et al. (1986) used analytic models and dissipative
%particle collisions in N-body simulations to study the response of the dark matter halo to the presence of baryons. More recently, other authors (e.g., Gnedin
%et al. 2004, and Sellwood et al. 2005) improved Blumenthal's model. 
%
%In Sect. 3.1, we describe how the baryon fraction, namely the ratio between the baryon mass and total mass was calculated and its adopted values.
%
%The random component
%of the CR introduces density perturbations on
%all scales, thus leading to major mergers (Romano-D´iaz
%et al. 2006, 2007). The mass inside the computational
%sphere is  6.1 × 1012 h-1M?. In the baryonic model,
%we have randomly replaced 1/6 of DM particles by equal
%mass SPH particles. Hence, m is not affected.

In the present paper, baryon fraction was calculated as follows. 

As discussed in the introduction, dwarf galaxies can be classified in galaxies having an high content of gas (e.g., dwarf irregular galaxies) and 
in galaxies having a low content of gas (e.g., dwarf elliptical galaxies or dwarf spheroidal).
Moreover, baryon content changes in the galaxy from the core to the outskirts (see Hoeft et al. 2006, Fig. 5).
The universal baryon fraction, $f_b$, has been inferred by observations involving different physical processes (Turner 2002). 
Percival et al. (2001), by using the Two Degree Field Galaxy Redshift Survey, obtained a value of $0.15 \pm 0.07$ , Jaffe et al. (2001) obtained $f_b= 0.186^{+0.010}_{-0.008}$ combining several data sets, while by using WMAP collaboration data (Spergel et al. 2003) ($\Omega_b=0.047 \pm 0.006$; $\Omega_m=0.29 \pm 0.07$), one obtains $f_b \simeq 0.16$. However, the content of baryons in galaxies is usually much smaller than $f_b$, since after baryons cool and form stars, the baryon-to-total mass at the virial radius 
is no longer equal the previous quoted universal baryon fraction and may deviate from it depending on feedback effects, hierarchical formation details, and heating by the extragalactic UVB flux. One has to take into account that for large galaxies, only about half of the baryons, in principle available within the virial radius, goes into the central galaxy. For dwarf galaxies, this amount may be much smaller because of feed-back effects, hierarchical formation details, and heating by the extragalactic UVB flux.
Geha et al. (2006) studied the baryon content of low mass dwarf galaxies and Hoeft et al. (2006) studied structure formation in cosmological void regions investigating to which extent the cosmological UV-background reduces the baryon content of dwarf galaxies. Hoeft \& Gottl\"ober (2010) studied also the case of cosmological environment. 
As shown from McGaugh et al. (2010) (see their Fig. 2), baryon fraction deviates from the cosmic baryon fraction:
while on dwarf scale only 1\% of the expected baryons are detected, on cluster scale most of them are detected. 
The detected baryon fraction is the ratio of the baryon fraction, $F_b=M_b/M_{500}$, and the universal baryon fraction, $f_b$: 
\begin{equation}
f_d = (M_b/M_{500})/f_b=F_b/f_b
\end{equation}
where $M_{500}$ is the mass contained within a radius where the enclosed density is 500 times the critical density of the Universe, and the baryon mass $M_b=M_{\ast}+M_{gas}$, where $M_{\ast}$ is the stellar mass and $M_{gas}$ is that of gas. For example for a dwarf of $M_b=2 \times 10^{7} M_{\odot}$, with $M_{500}= 2 \times 10^{9} M_{\odot}$ (see Table 2 in McGaugh et al. 2010), $f_d$ is $\simeq 0.04$, and the baryon fraction $F_b=M_b/M_{500}=0.01$.

The result of our model (extensively described in DP09) shows that the structure of the inner density profiles is strongly influenced
by baryon physics and exchange of angular momentum between baryons and dark matter.
So, summarizing, in our analytical model the density profiles are: \\
a) core--like, in the inner part of the density profile, when we take into account the presence of baryons. In the outer parts the result is the same of simulations. \\
b) The results of dissipationless simulations, namely cuspy profiles, are re-obtained, with very good accuracy, eliminating the baryons, not taken into account in dissipationless simulations. The results are in agreement with other models, e.g. Mashchenko et al (2005, 2006, 2007),
Romano-Diaz et al. (2008), El-Zant et al. (2001), and G10.

As reported in the introduction and also in this section, the model of the present paper has some peculiar features 
that makes it unique: a) the other semi-analytical models take into account only random angular momentum, while the present paper takes also into account ordered angular momentum. Other studies, concentrate only on the study of one effect: angular momentum or dynamical friction or adiabatic contraction, while we
study their joint effect. b) We can take into account sub-grid physics accurately and without using too high CPU time, and this allow us to study how change of angular momentum or baryon fraction influence density profiles.

%\subsection{Baryon fraction}

\section{Results and discussion}

\subsection{Density profiles, torque and baryons fraction}

While very successful on large scale, $\Lambda$CDM model suffers from a number of problems on galactic scales (Flores \& Primack 1994; Moore 1994; Ostriker \& Steinhardt 2003). 
One of these, the cusp-core problem is particularly troublesome and despite many efforts it has not entirely been resolved to date. 
In DP09, we showed how the flattening of the inner slopes of halos is produced by the role of angular momentum, dynamical friction, and the interplay between DM and baryon component. In peculiar, we studied the density profiles of haloes in the mass range $10^{8}-10^{15} M_{\odot}$. 
%In the case of the dwarf galaxies mass range, the density profiles were core-like.
As previously reported, in that study we moreover took into account the ordered angular momentum, due to tidal torques (see Section 2 of the present paper and Section C of DP09), experienced by proto-halos in the neighborhood of the studied object. It would be then interesting to study the effects of different angular momenta and baryon fraction on the density profile of a given halo.
So, in the following, using DP09, we will generate density profiles of haloes in the mass range $10^{8}-10^{10} M_{\odot}$ and we will study the effect of different ordered angular momentum and baryon fraction on the dwarf galaxy density profile. 
In a future paper, we also aim to study observationally the inner slopes of dwarfs who has no interacting companions, and those more close to parent galaxies. 
In Sect. 3.1, we use the baryons fraction of McGaugh et al. (2010). In Sect. 3.2, where we compare the results with three observed galaxies, we will discuss how the baryon fraction is chosen, for those peculiar cases.
The result of the calculation are shown in Fig. 1a, in which we plotted the density profiles of dwarfs with mass $10^{8} M_{\odot}$ (dashed line), 
$10^{9} M_{\odot}$ (dotted line), and $10^{10} M_{\odot}$ (solid line). This panel shows the results for haloes density profiles obtained in DP09 (which we will use as reference haloes). The angular momentum in Fig. 1a was obtained by the tidal torque theory as in DP09 and, in the case of the halo of $10^9 M_{\odot}$ has a value similar to UGC 7399 (see BBS), namely $h_{\ast} \simeq 400$ kpc km/s ($\lambda \simeq 0.04$), where $h$ is the specific ordered angular momentum. The baryon fraction ($f_{d_{\ast}} \simeq 0.04$)\footnote{Note that the ``$\ast$" in $f_{d_{\ast}}$ and in $h_{\ast}$ does not ``stand" for stellar, as in the case of $M_{\ast}$} 
was fixed as described Sect. 2.
The quoted plot shows flat density profiles well fitted by a Burkert profile, whose functional form is characterized by:
\begin{equation}
\rho(r)= \frac{\rho_o}{(1+r/r_o)[1+(r/r_o)^2]}
\end{equation}
where $\rho_o \simeq \rho_s$ and $r_o \simeq r_s$ (El-Zant et al. 2001).

\begin{figure*}
\centering
(a)
\hspace{0.8cm}
\subfigure{\includegraphics[width=6.4cm]{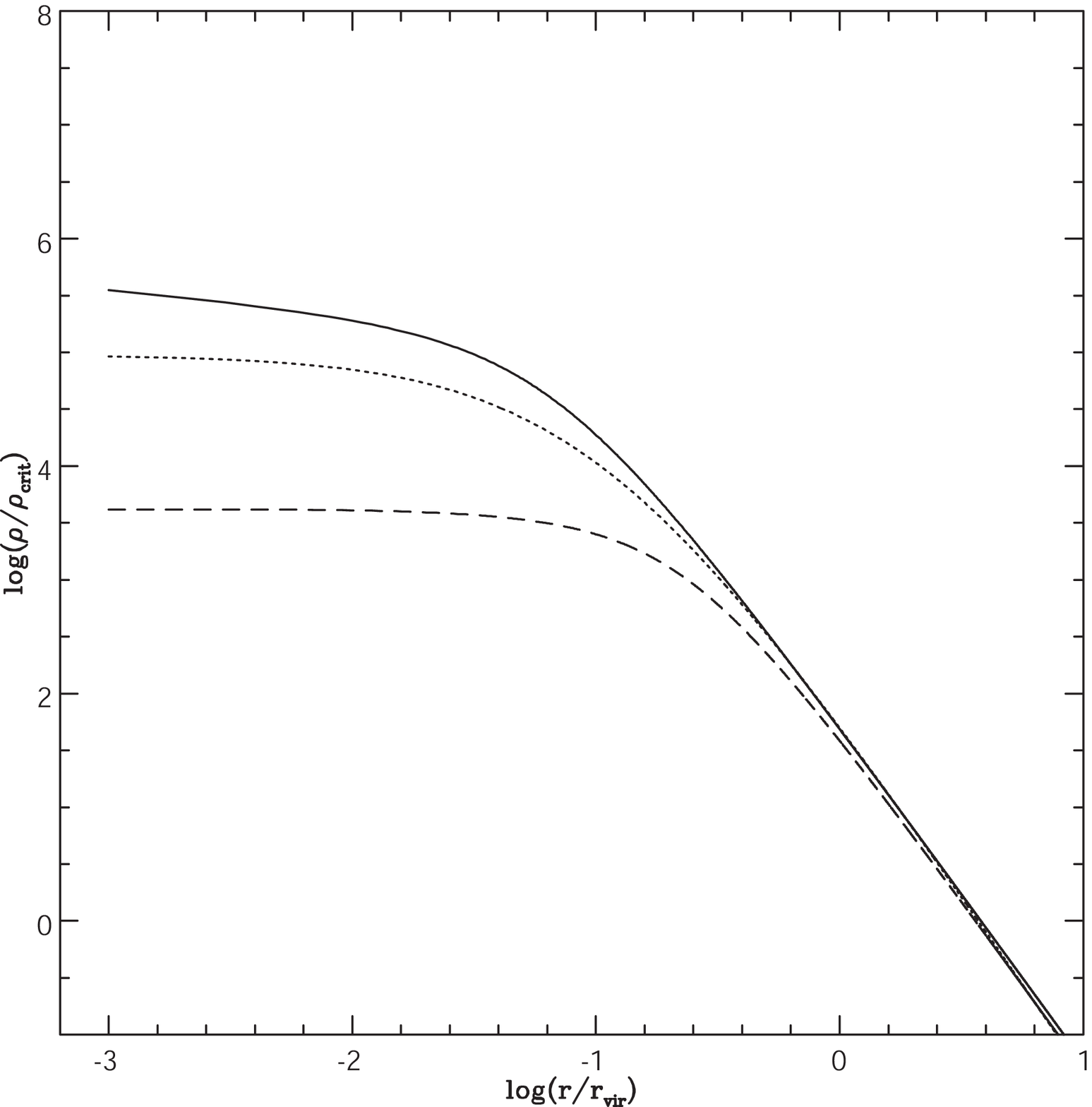}} (b) \goodgap 
%\includegraphics[width=84mm]{dwarfsnmm.eps}
%\hspace{-0.15cm}
\subfigure{\includegraphics[width=6.6cm]{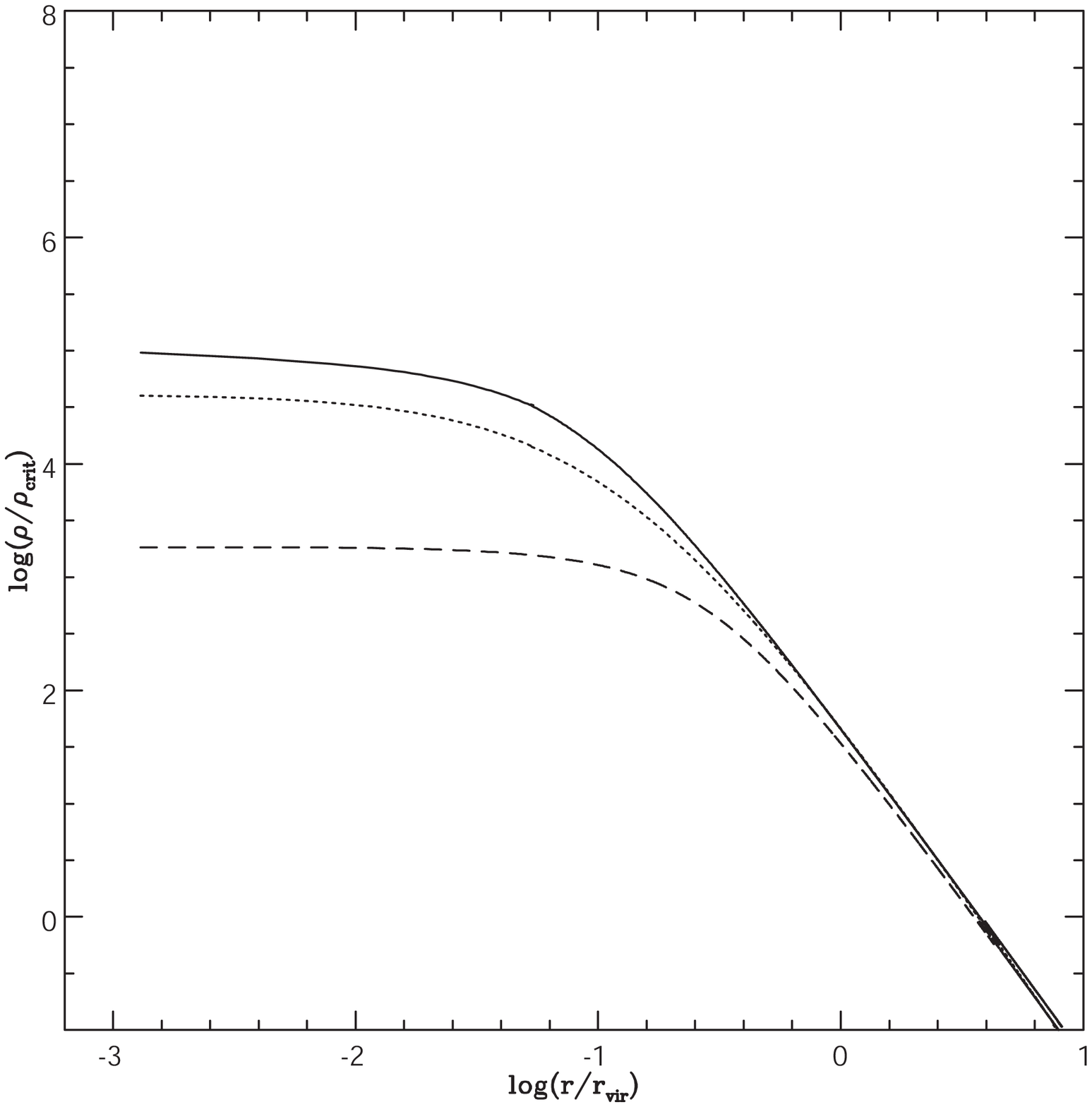}}  \\ (c)\goodgap 
%\hspace{-0.5cm}
\subfigure{\includegraphics[width=6.8cm]{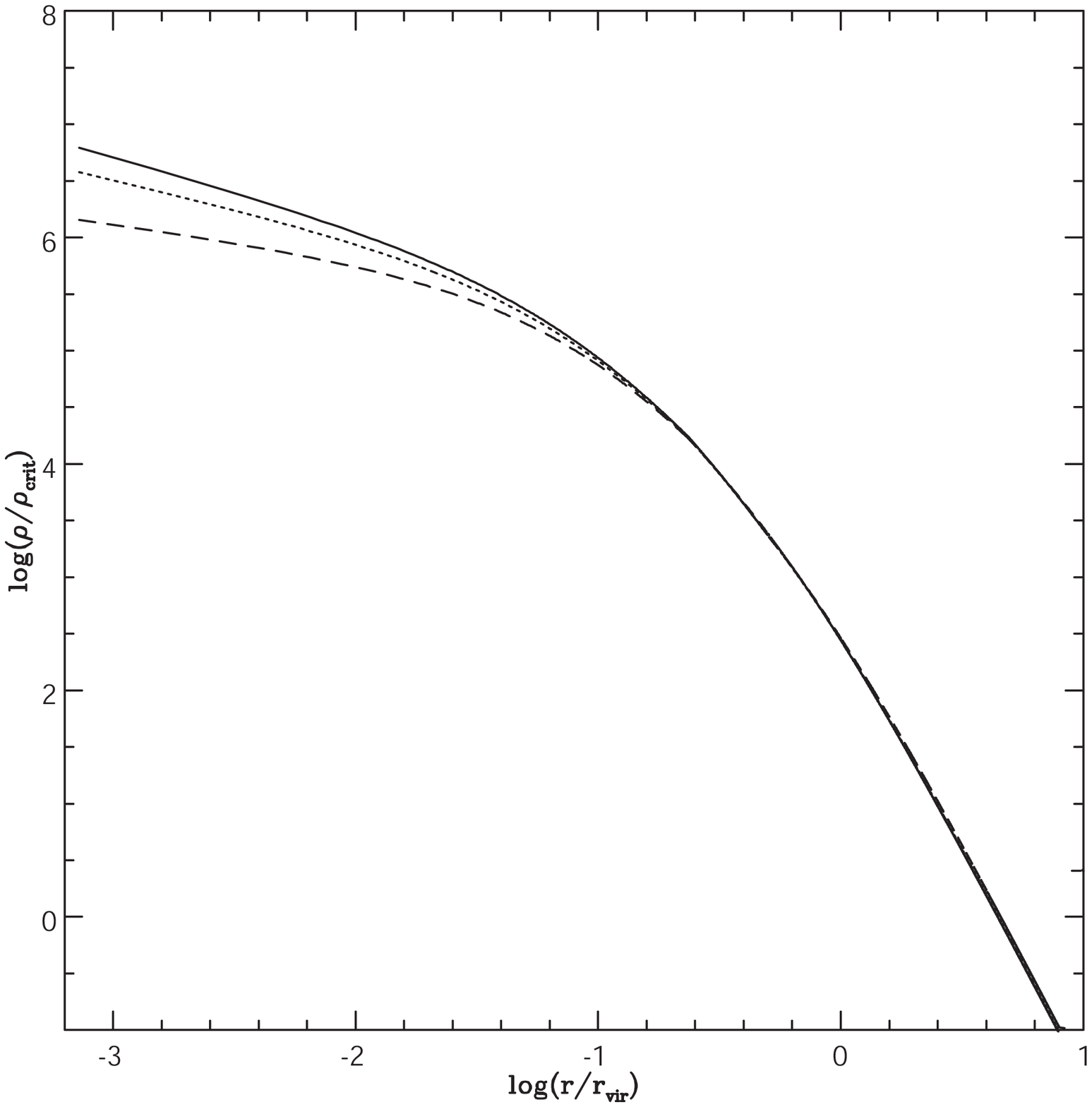}} (d)  \goodgap  
%\includegraphics[width=84mm]{dwarfs33.eps}
%\hspace{0.15cm} \includegraphics[width=68mm]{dwarfs4i.eps}
%\hspace{-0.15cm} 
\subfigure{\includegraphics[width=6cm]{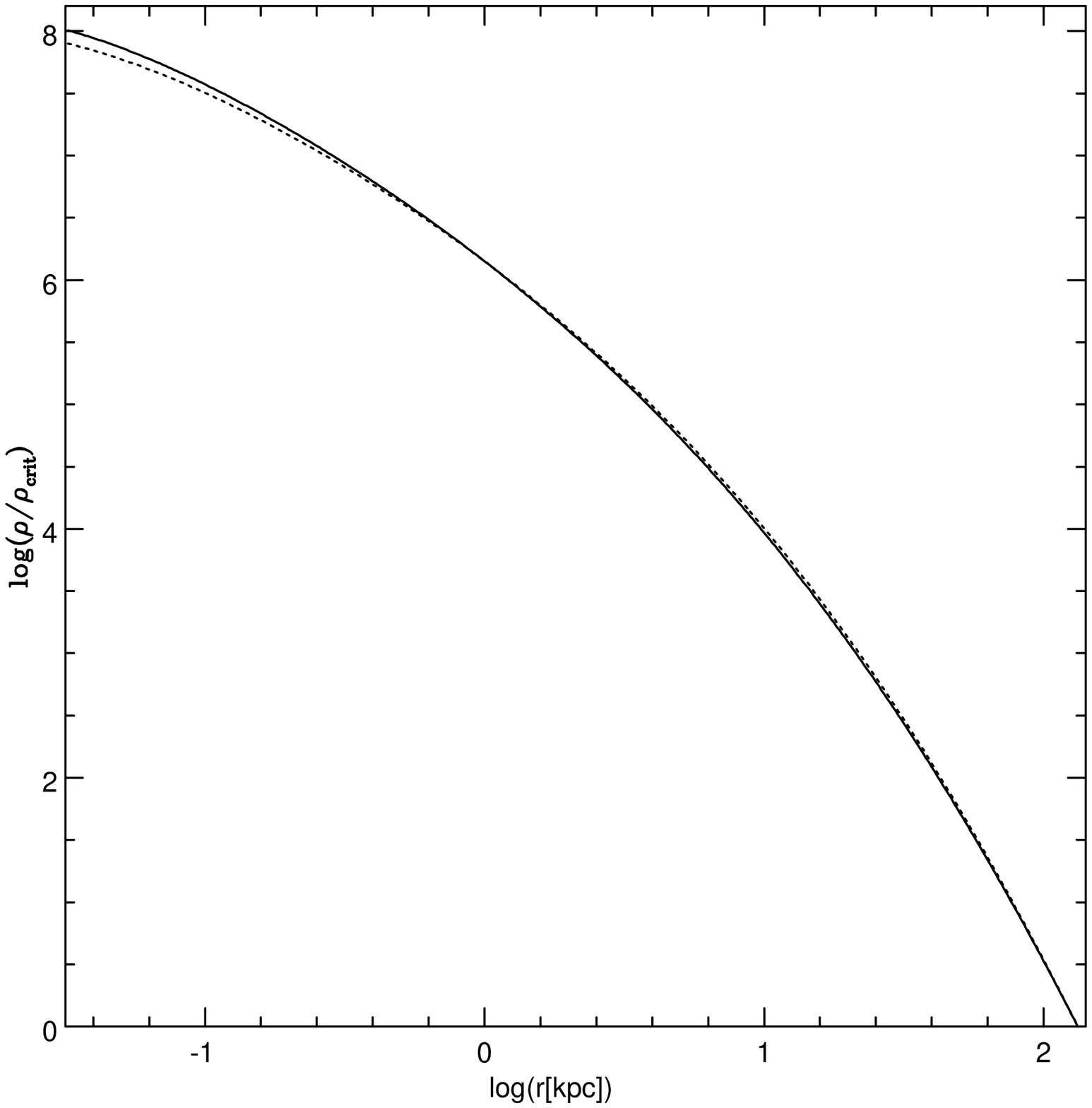}} \goodgap \\
%\hspace{0.15cm} \includegraphics[width=68mm]{dwarfs4ii.eps}
%\includegraphics[width=84mm]{dwarfsn_isol.eps}
\caption{Shape changes of dark matter haloes with angular momentum.
%with angular momentum. DM haloes generated with the model of Section 2. 
In panels (a)--(c), the dashed line, the dotted line and the solid line represent the density profile for a halo of $10^8 M_{\odot}$, $10^9 M_{\odot}$, and $10^{10} M_{\odot}$, respectively. %Halos in panels (a)-(d) are obtained by means of Del Popolo (2009). 
In case (a), that is our reference case, the specific angular momentum was obtained using the tidal torque theory as described in DP09. The specific angular momentum for the halo of mass $10^9 M_{\odot}$ is $h_{\ast} \simeq 400$ kpc km/s ($\lambda \simeq 0.04$) and the baryon fraction 
$f_{d_{\ast}} \simeq 0.04$. 
%we used the same parameters of the model, while 
In panel (b) we increased the value of specific ordered angular momentum, $h_{\ast}$, to $2 h_{\ast}$ leaving unmodified the baryon fraction to $f_{d_{\ast}}$ and in panel (c) the specific ordered angular momentum is $h_{\ast}/2$ and the baryon fraction equal to the previous cases, namely $f_{d_{\ast}}$.
%equal to zero (isolated system).
Panel (d) shows the density profile of a halo of $10^{10} M_{\odot}$ with zero ordered angular momentum and no baryons (solid line), while the dashed line is the Einasto profile.}
\end{figure*}

\begin{figure*}
%\hspace{-4cm} 
\centering
(a)
\hspace{0.2cm} 
\subfigure{\includegraphics[width=6.4cm]{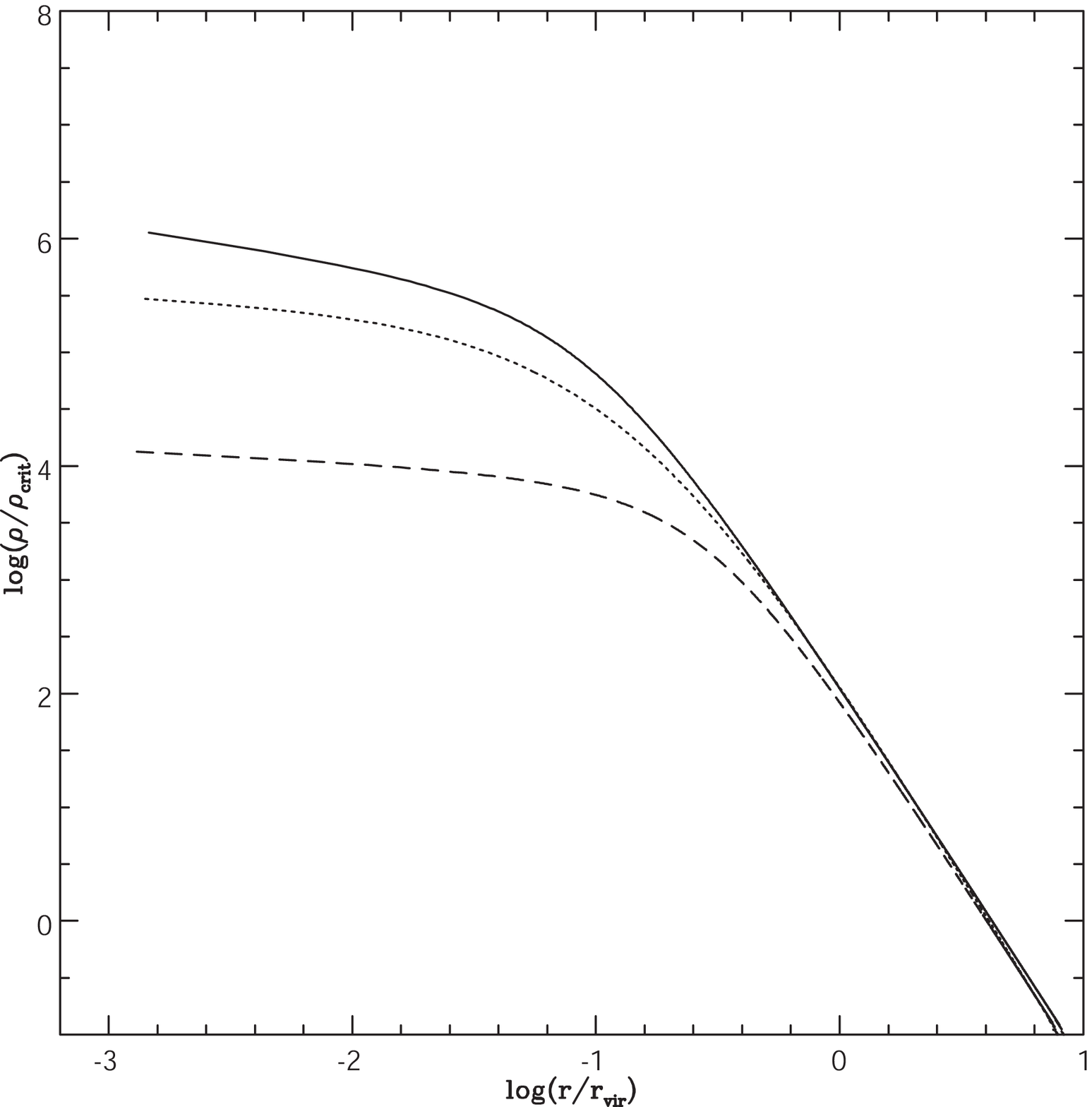}} (b) \goodgap 
\subfigure{\includegraphics[width=6.8cm,height=5.4cm]{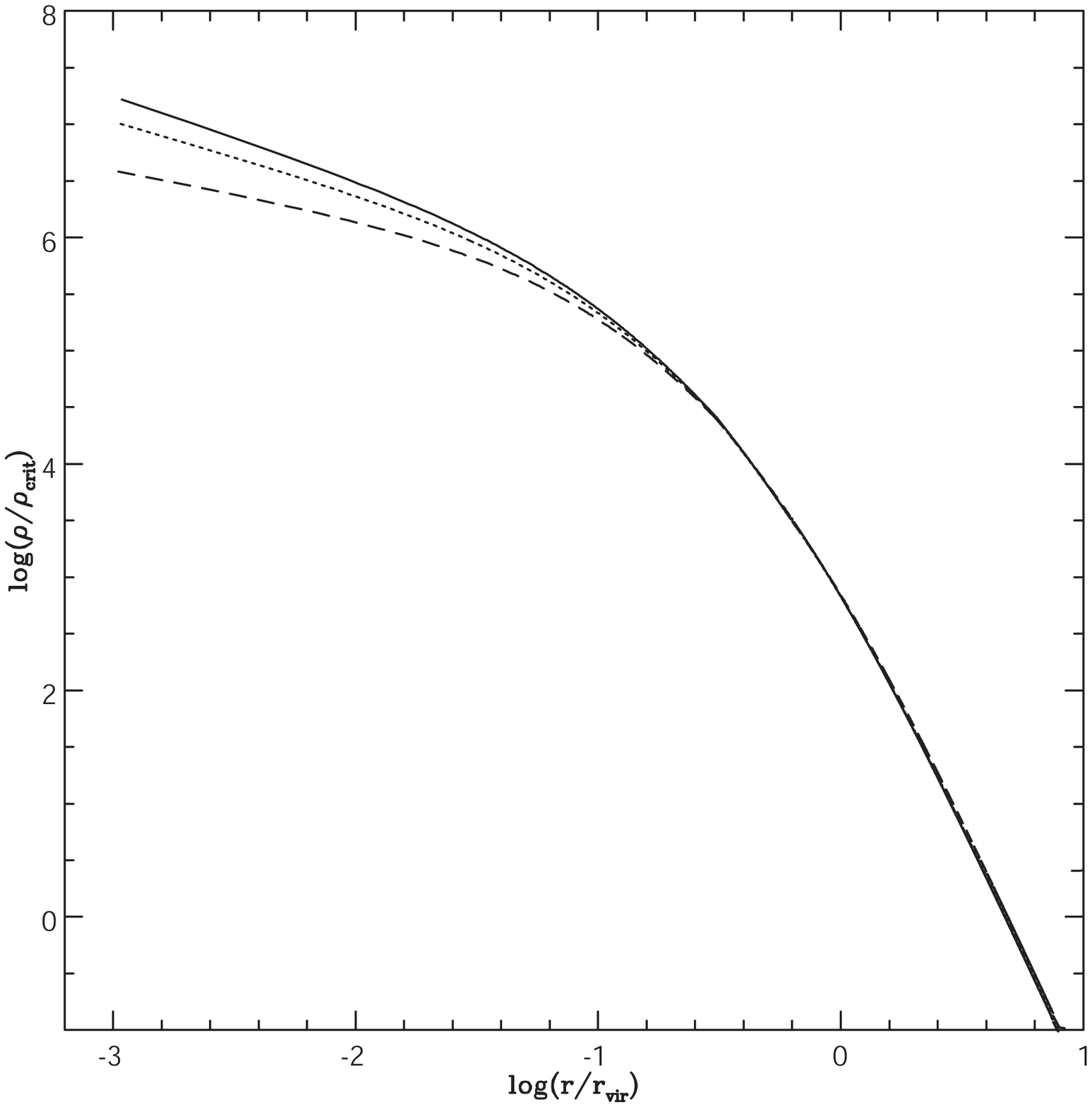}} \\ (c) \goodgap
\subfigure{\includegraphics[width=6.8cm]{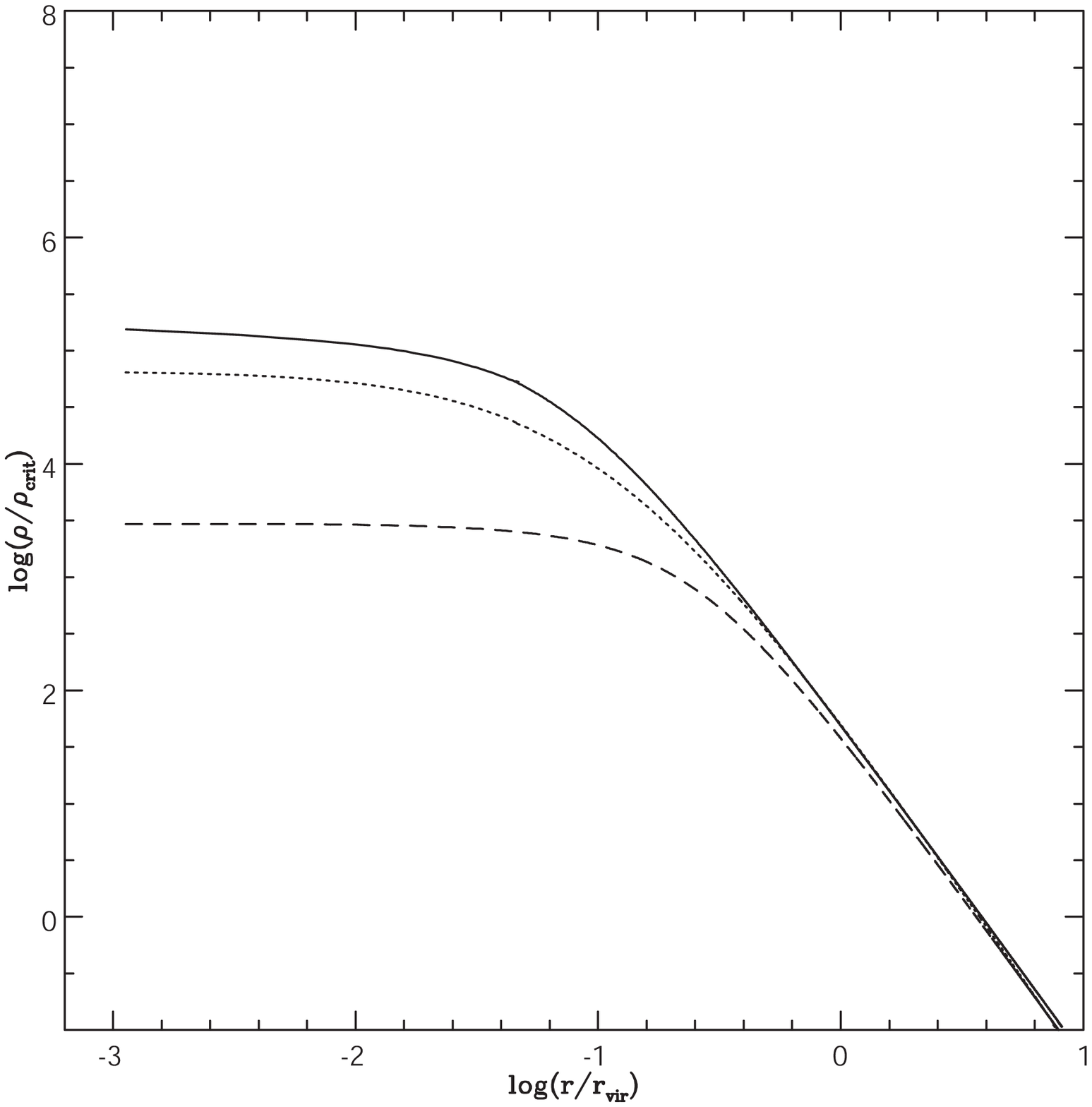}} (d) \goodgap
\subfigure{\includegraphics[width=6.8cm]{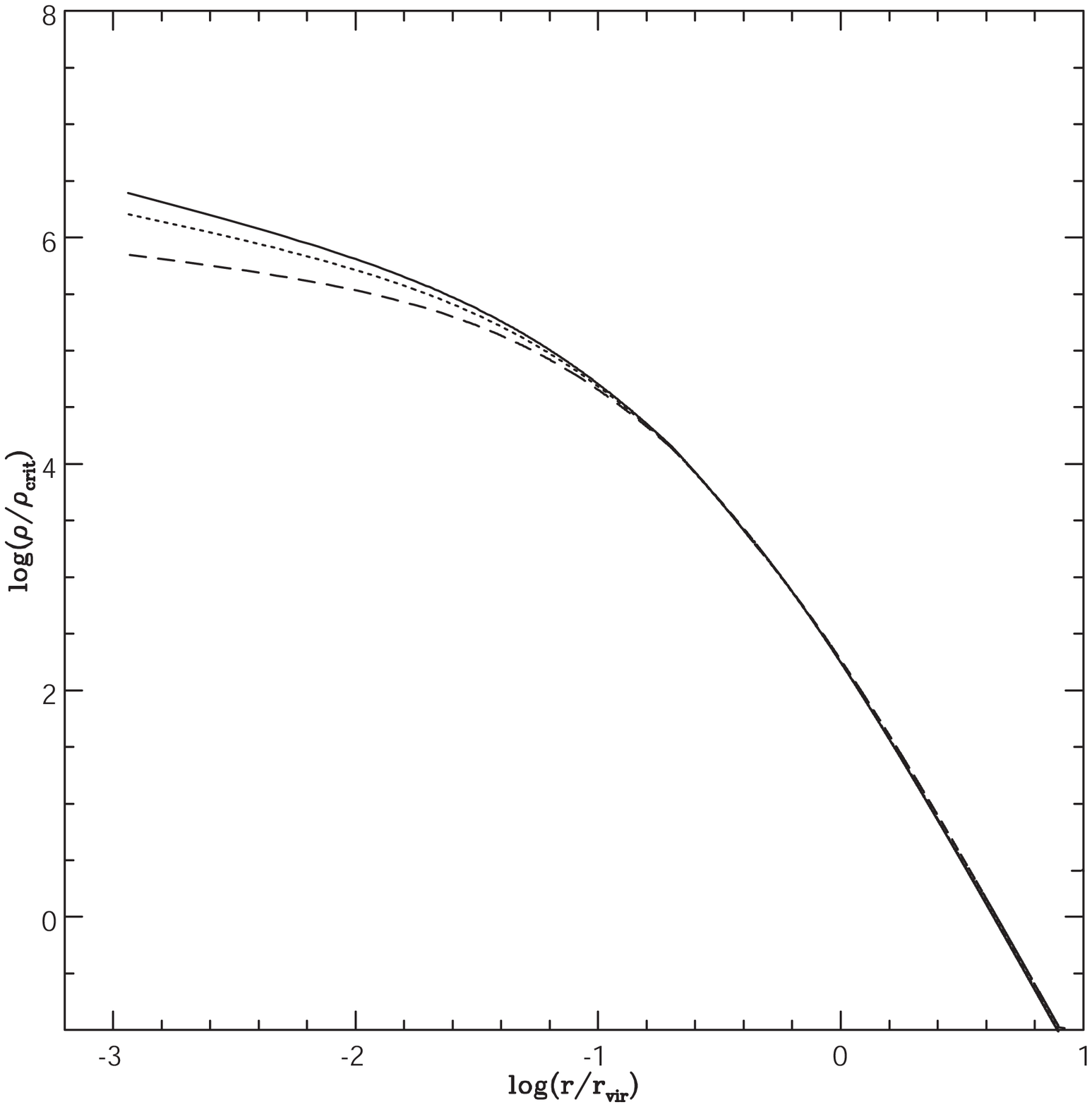}}  \goodgap \\
\caption{DM halos shape changes with baryon fraction. Same as previous figure, Fig. 1, but in panel (a) we reduced the value of baryon fraction of Fig. 1a ($h_{\ast}$, $f_{d_{\ast}}$) to $f_{d_{\ast}}/3$, and in panel (b) we reduced the value of baryon fraction of Fig. 1c to $f_{d_{\ast}}/3$; in panel (c) we increased the value of baryon fraction of Fig. 1a ($h_{\ast}/2$, $f_{d_{\ast}}$) to $3 f_{d_{\ast}}$, and in panel (d) we increased the value of baryon fraction of Fig. 1c to $3 f_{d_{\ast}}$.
%but now the value of specific ordered angular momentum and baryon fraction are: $h_{\ast}$, $F_b_{\ast}/2$ (panel a); 
%$h_{\ast}/2$, $F_b_{\ast}/2$ (panel b); $h_{\ast}$, $2 F_b_{\ast}$ (panel c); $h_{\ast}/2$, $2 F_b_{\ast}$ (panel (d)  
%
%we changed the baryon fraction to of Fig. 1a of a factor 1/2; in panel (b) we did the same for panel (c) of Fig. 1. Panel (c) and (d) are similar to panel (a) and (b) but %now we changed the baryon fraction of factor 1/2
}
\end{figure*}

 A slight steepening of the density profile is visible in the case $10^{10} M_{\odot}$ in which case the inner slope of the profile is $\alpha \simeq 0.2$, while $\alpha \simeq 0$ for $10^{8} M_{\odot}$-$10^{9} M_{\odot}$. 

As previous discussed in Sect. 2, the ordered angular momentum acquired by proto-structures depends on mass
and is higher for dwarf galaxies, moreover the values of $\lambda$ has a range $\simeq 0.02-0.2$ (BBS; Boisier et al. 2003). So, after calculating the density profiles for value given by the tidal torque theory (TTT), we considered the effects of changing the values of ordered angular momentum in the range of the allowed values. 
In Fig. 1b, we repeated the same calculation of panel (a) changing the amplitude of the ordered angular momentum (which is proportional to the tidal torque) to 
$2 h_{\ast}$, leaving unchanged the baryon fraction to $f_{d_{\ast}}$.
In this case, the profiles are, as expected, flatter with respect to the reference case, and for the same mass as the reference case ($10^{8} M_{\odot}$, $10^{9} M_{\odot}$, $10^{10} M_{\odot}$) the slope is $\alpha=0$, in all three cases.
In Fig. 1c, we plot the density profiles for haloes having a value of angular momentum $h_{\ast}/2$, ($1/2$ that of the reference case), leaving unchanged the value of baryon fraction to $f_{d_{\ast}}$. 
In this case, the inner slope for haloes of the same mass ($10^{8} M_{\odot}$, $10^{9} M_{\odot}$, $10^{10} M_{\odot}$) are, respectively, given by $\alpha \simeq 0.4$, 
$\alpha \simeq 0.5$, and $\alpha \simeq 0.6$. The previous results show a steepening of the slope with decreasing value of angular momentum, in agreement with several previous results (Sikivie et al. 1997; Avila-Reese et al. 1998; Nusser 2001; Hiotelis 2002; Le Delliou \& Henriksen 2003; Ascasibar et al. 2004; Williams et al. 2004).  
The three panels show that: a) the less massive is a halo the less concentrated it is; b) larger mass, and higher angular momentum haloes are characterized by smaller inner slopes.\\
Point (a) can be explained as follows: higher peaks (larger $\nu$), which are progenitors of more massive haloes (Peacock \& Heavens 1990; Del Popolo \& Gambera 1996, Gao \& White 2007) have greater density contrast at their center, and so shells do not expand far before beginning to collapse. This reduces the angular momentum acquired and allows haloes to become more concentrated.
Point (b) can be explained as folllows. Less massive objects are born from peaks with smaller height, $\nu$, and they acquire more ordered, $h$, and random, $j$, angular momenta. The density profile shape in the inner regions of haloes are particularly sensitive to angular momentum.  In case of pure radial orbits, the core is dominated by particles from the outer shells. The larger is the angular momentum of particles, the 
longer they remains closer to the maximum radius. This give rise to a shallower density profile. Particles with smaller angular momentum will enter the core with a smaller radial velocity, with respect to particles in a pure radial SIM. 
Some particles have an angular momentum so large to be unable to fall into the core. So, particle with larger angular momentum will not reach the halo center and will not contribute to the central density, and this gives rise to flatter density profiles. 
Summarizing, larger ordered angular momentum (stronger interaction between the proto-dwarf and environment) gives rise to flatter density profiles, while a larger value of mass produces an increase of the central density contrast, and a steeper profile. 

Fig. 1d shows the case of zero angular momentum and no baryons, for a halo of $10^{10} M_{\odot}$. The solid line is the result of the model of the present paper, while the dashed line is the Einasto profile, characterized by a logarithmic slope varying continuously with radius (Navarro et al. 2004, 2008), and having a functional form:
\begin{equation}
\ln{(\rho/\rho_{-2})}=(-2/\eta) [(r/r_{-2})^{\eta} -1]
\end{equation}  
where the characteristic radius, $r_{-2}$, is obtained by $d \ln{\rho}/d \log{r}=-2$, $\rho_{-2} \equiv \rho(r_{-2})$, and $\eta$ is a free parameter, which for $\eta \sim 0.17$ reproduces fairly well the radial dependence of density profiles. In Fig. 1d, we did not add the plot of NFW profile since recent simulations (e.g., Navarro et al. 2004, Navarro et al. 2010) shows that density profiles from N-body simulations are better approximated by Einasto profiles. 
The correlation between steepness of density profiles and angular momentum is very important to have a better understanding of 
LSB galaxies. We recall that LSB galaxies are more angular-momentum-dominated than normal galaxies of the same luminosity (McGaugh \& de Blok 1998; Boisier at al. 2003), implying a flatter density profile with respect to giant galaxies.
Another parameter which influence the final density profile is the baryon content of the halo, namely the ratio of the baryon mass to the total mass. 
In the following, we will study how changes in baryon content affects the density profiles. 
To this aim, in Fig. 2a, we plotted the same density profiles of Fig. 1a 
but with a baryon fraction $f_{d_{\ast}}/3$, and the value of specific angular momentum as left unchanged to $h_{\ast}$. The effect of reducing the baryon content has the effect of steepening the density profiles, with respect to the reference case. The inner slope for haloes having the same mass as the reference case 
($10^{8} M_{\odot}$, $10^{9} M_{\odot}$, $10^{10} M_{\odot}$) are, respectively, given by $\alpha \simeq 0.13$, $\alpha \simeq 0.16$, and $\alpha \simeq 0.32$. Fig. 2b represents the haloes of Fig. 1c ($h_{\ast}/2$, $f_{d_{\ast}}$) but reducing the baryon fraction to $f_{d_{\ast}}/3$. The inner slope are steeper than the one seen in Fig. 1c, namely we have $\alpha \simeq 0.4$, $\alpha \simeq 0.6$, and $\alpha \simeq 0.7$. Fig. 2c represents the same density profiles in Fig. 1a ($h_{\ast}$, $f_{d_{\ast}}$)
but in this case the baryon fraction is larger, namely $3 f_{d_{\ast}}$. Fig. 2d represents the same density profiles in Fig. 1c ($h_{\ast}/2$, $f_{d_{\ast}}$)
but in this case the baryon fraction is larger, namely $3 f_{d_{\ast}}$.
A larger amount of baryons imply a flattening of the density profiles. Fig. 2c shows density profiles with $\alpha \simeq 0$, for the three previous mass values, 
and Fig. 2d shows density profiles with $\alpha \simeq 0.3$, $\alpha \simeq 0.45$, and $\alpha \simeq 0.51$. 
The tendency to have flatter profiles with increasing baryon content and shallower with larger baryon content is due 
to the fact that when more baryons are present the energy and angular momentum transfer from baryons to DM is larger and DM moves on larger orbits reducing the inner density. 

In Fig. 3, we check the model against G10 SPH simulations of dwarf galaxies in a $\Lambda$CDM framework.
The figure was obtained following the prescription of G10 to calculate density profiles. 
In SPH simulations, G10 studied the formation of dwarf galaxies in a $\Lambda$CDM cosmology including baryonic processes, as gas cooling, heating from the cosmic UV field, star formation and SN driven gas heating was included. Dark matter particle masses in the high resolution regions had $1.6 \times 10^4 M_{\odot}$, and the force resolution, i.e., the gravitational softening, is 86 pc.
They studied two  dwarf galaxies, with mass $3.5$ (DG1) and $2.0 \times 10^{10} M_{\odot}$ (DG2) in a $\Lambda$CDM cosmology with $\sigma_8 = 0.77$, $\Omega_M = 0.24$, $\Omega_{\Lambda}= 0.76$, $h = 0.73$, and $\Omega_b = 0.042$.

%We calculated the density profile using Governato's prescriptions. 
%
%In peculiar in the first case, we did not include baryons, since Stadel's simulations are dissipationless.  
%In Fig. 2a, we compare the Stadel's simulation result (solid line) with our model's result (dashed line). The plot shows a good agreement between our result and that of %Stadel. 
Fig. 3 plots Governato's DG1 (solid blue line) and DG2 (solid black line) density profiles. Our profile is the dot dashed line. The plot shows a good agreement of our density profile with SPH results. 

\begin{figure*}
\centering
\subfigure{\includegraphics[width=10.4cm]{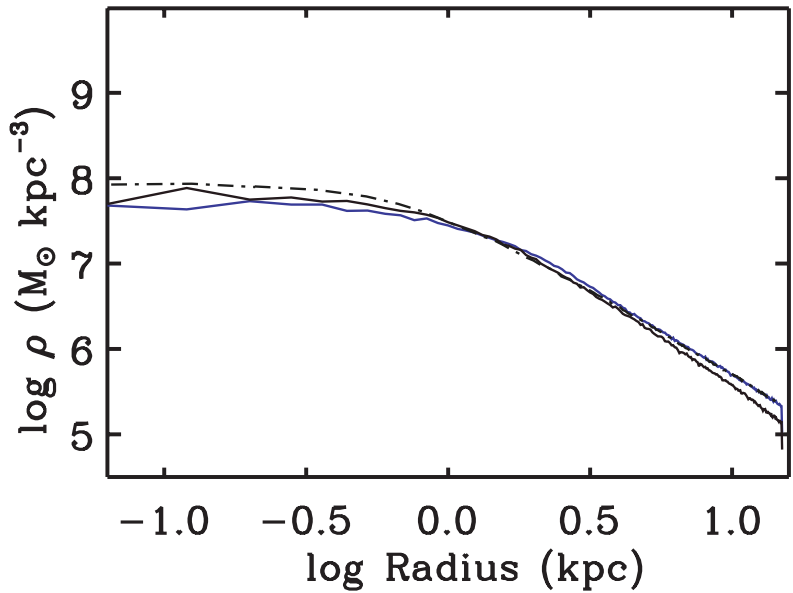}}  \goodgap
\caption{Comparison of the dark matter density profile of our model with simulations. Figure plots the result of our model (dot-dashed line) and galaxy DG1 (solid blue line) and galaxy DG2 (solid black line) of G10 SPH simulations.
}
\end{figure*}

\subsection{Comparison with observed galaxies}

In the following, we use mass models for the dark matter component of different dwarf galaxies, from a dwarf with flat inner profile (NGC 2976), to an intermediate one 
(NGC 5949) to one with a steep profile (NGC 5963) and compare these with those obtained from the model of Sect. 2.

\subsubsection{NGC 2976}

In Figs. 4a-b, we apply the model of Sect. 2 to find density profiles, as done in sect. 3.1, that we then  
compare to the minimum disk and maximum disk rotation curve of NGC 2976, respectively. 
Before, we summarize the characteristics of the quoted dwarf fully described in Simon et al. (2003) (hereafter S03). 
NGC 2976 is a regular Sc dwarf galaxy, a pure disk system, unbarred, and bulgeless (see Fig. 1 of S03). The dwarf, located in the group of M81 has a total mass, at 2.2 kpc, of  $3.5 \times 10^9 M_{\odot}$.
% and an absolute magnitudes of $M_B=-17.0$. 
S03 obtained, for the stellar disk, average values of $0.48 \pm 0.02 M_{\odot}/L_{\odot K}$ in K band and $1.07 \pm 0.07 M_{\odot}/L_{\odot R}$ in R band. Stellar disk and gaseous disk constitute the fundamental reservoir of baryons of the dwarf. NGC 2976 has also a molecular disk and a gas disk, whose mass was estimated to be $1.5 \times 10^8 M_{\odot}$, by Appleton, Davies, \& Stephenson (1981), and Stil \& Israel 2002.   
The kinematics of the gaseous disk is highly complex, exhibiting large non-circular motions near the center which could be he reflection of a triaxial halo (Kazantzidis et al. 2010). S03 determined, for NGC 2976, the minimum and maximum disk rotation curve, represented by the black circles with error-bars in Fig. 4a, 4b, respectively.
S03 also fitted the rotation curve with a power-law from the center to $r \simeq 110"$ (1.84 kpc) (solid line in Fig. 4a) and they found that the corresponding 
density profile, representing the density corresponding to the total mass (baryon plus dark matter), is well fitted by $\rho_{tot}=1.6 (r/1 pc)^{-0.27 \pm 0.09} M_{\odot}/pc^3$. In Fig. 4a, the dashed line represents the result of the model of section 2 of the present paper considering the total mass (sum of dark matter and baryons) (note that this is the only case in the paper where we plot he total mass: dark matter and baryons in order to compare results with the minimum disc rotation curve of S03), using $\lambda \simeq 0.04$, compatible with the value of $\lambda$ given in Mun$\tilde{o}$z-Mateos et a. (2011). 
%The baryon fraction was calculated as follows.
The baryon fraction in the case of NGC 2976 could be obtained as $(M_{gas}+M_{\ast})/M_{tot}$, where $M_{tot}=3.7 \times 10^9 M_{\odot}$ (S03), $M_{gas}=1.5 \times 10^8 M_{\odot}$ (5\% of the total mass) and $M_{\ast} = 5 \times 10^8 M_{\odot}$ (14\% of the total mass) (SO3). However, the values of the mass given refers, as previously reported, only to the inner $2.2$ kpc of the dwarf, so the quoted ratio would give just the baryon fraction of the inner part of the dwarf. For NGC 2976 as also for NGC 5949, and NGC 5963 there is not enough data to determine the total mass and then the baryon fraction, 
since the dark matter halo is vastly more extended than any of the baryon tracers that are needed to measure the mass. So, one have to make a series of very broad assumptions, as done in McGaugh et al. (2010).
In the case of NGC 2976, we know the inner masses and the circular velocity $V_c \simeq 75$ km/s. Governato et al. (2007) simulated a disk galaxy with $V_c = 70$ km/s 
and found that it has a virial mass of $1.6 \times 10^{11} M_{\odot}$, similarly to Kaufmann et al. (2007), who simulated galaxies with $V_c = 74$ km/s getting $M_{vir} = 1.1 \times 10^{11} M_{\odot}$). 
So, as for Fig. 1, 2, we shall use McGaugh estimates, taking account of the known rotation circular velocity of NGC 2976 and the simulated mass in Governato et al. (2007). This gives a value $f_d \simeq 0.1$. 
The plot shows that, on the entire length of the data, the model of this paper gives a very good fit to the rotation curve, better than the power-law fit obtained by S03 or a NFW fit (here not plotted). The shape of the density profile of the dark matter halo (Fig. 4b), was obtained by S03 removing the rotational velocities contributed by the baryon components of the galaxy, 
%As first step, the stellar disk is scaled as high as the observed rotation velocities allow (see also Fig. 10 of S03), obtaining a maximum disk having a mass-to-light ratio %of $0.19 M_{\odot}/L_{\odot K}$. 
%Then subtracting the rotation velocities due to the stars, the rotation velocities due to the HI, and the rotation velocities due to the H2 in quadrature from the observed %rotation curve, one obtains 
obtaining the dark matter density profile of the maximal disk, 
%\begin{equation}
$\rho_{DM}=0.1 (r/1pc)^{-0.01 \pm 0.13} M_{\odot}/pc^3$
%\label{eq:dm}
%\end{equation}
In Fig. 4b, the black circles with error-bars represent the dark matter rotation velocities for the maximal disk, as previously reported, and the solid line  
is the power-law fit 
%to the halo velocities 
(for $14" < r < 105"$) obtained by S03. 
Now, from the previous discussion, the slope of the total density profile of the galaxy ($\alpha_{DM} \leq 0.27 \pm 0.09$), composed of dark matter stars and gas, 
represents the absolute upper limit for the slope of the dark matter density profile. The lower limit to the dark matter density profile, 
obtained from the maximal disk, is $\alpha_{DM} \leq 0.01 \pm 0.13$.
%This profile is composed of dark matter stars and gas, so the upper limit must be lower. 
S03 concluded, after excluding strong and very young starburst, that the dark matter density profile is bracketed by $\rho_{DM}=r^{-0.17 \pm 0.09}$ and $\rho_{DM}=r^0$. 
In Fig. 4b, the dashed line represents again the result of the model of section 2 of the present paper, now considering only dark matter. 
Also in this case, the model of this paper gives a very good fit to the rotation curve, better than NFW model and the power-law fit. 

In the previous discussion, we saw that the density profile of NGC 2976 is very shallow, $\rho_{DM}=0.1 (r/1pc)^{-0.01 \pm 0.13} M_{\odot}/pc^3$, compatible with some results in literature (e.g., de Blok et al. 2003), but in contradiction with the N-body simulations results (Stadel et al. 2009 found a minimum value of  the slope $\alpha=-0.8$ at 120 pc). As discussed by several papers (e.g. DP09; Governato et al. 2010), the shallow central density profile of NGC 2976 does not necessarily imply a problem for CDM model. Here, we only recall that one reason for the discrepancy between simulations and observations is due to the fact that  
simulations neglect the effects of the baryons on the dark matter halo. 
In the peculiar case of NGC 2976, baryons dominate the central region of the galaxy out to 220 pc, for the lower limit to mass-to-light ratio, and 
out to a radius of 550 pc, in the case of the maximal disk (S03). So, summarizing, the flat density profile of NGC 2976 is produced, as discussed in DP09,
by the role of angular momentum, dynamical friction, and the interplay between DM and baryon component. 
The importance of tidal torques in shaping dwarf galaxies density profile has been also suggested by several authors (e.g., Stoehr et al. 2002, and Hayashi et al. 2003). 
According to the previous works, smaller satellites of dark matter halos suffer tidal stripping, and tidal effects that mainly strip mass from the external regions of orbiting object and produce an expansion of the halo with the effect of a decrease in the central density at each peri-centric passage.
So, even if the effect of tidal field is more evident in the outer part of the dwarf, also the inner profiles is influenced and it is 
shallower than their original profiles. However, tidal stirring is more efficient in forming dwarf spheroidal (dSph) galaxies starting from progenitors that are late type dwarfs affected by tidal forces (Kazantzidis et al. 2010, Lokas et al. 2010, Sales et al. 2010).
The mechanism now quoted could flatten NGC 2976 density profile if it is member of M81 and if it is at such a distance that tidal effects of M81 can influence it.  
If we assume for M81 the total mass determined by Karachentsev et al. (2002), namely $10^{12} M_{\odot}$, in order that M81 tidal field is comparable to NGC 2976 gravity 
%(at a radius of 2 kpc), 
the two objects should approach each other at a distance of 20 kpc. Nowadays, their projected distance is known to be 79 kpc, and this means that, at present-time, NGC 2976 is probably unaffected by the interaction. This does not imply that if nowadays NGC 2976 does not seem to be participating in the tidal interaction currently taking place between M81, M82, and NGC 3077 (Yun, Ho, \& Lo 1994), that it was not interacting with them in the past. 
The existence of a HI streamer connecting M81 and NGC 2976 (Appleton et al. 1981) is probably the rest of a past interaction. More recently, Boyce et al. (2001) showed that the gas  is composed of a single tidal bridge connecting the two galaxies (see their Fig. 2a).
%If NGC 2976 can be identified with one of the most massive few dark matter satellites of M81, the quoted mechanism would have some effects on the shallowing of the density %profile of the dwarf. Nowadays, NGC 2976 does not appear to be participating in the tidal interaction currently taking place between M81, M82, and NGC 3077 (Yun, Ho, \& Lo %1994). Assuming that M81 has a total mass of $10^{12} M_{\odot}$ (Karachentsev et al. 2002), its tidal field only becomes comparable to the gravity of NGC 2976
%(at a radius of 2 kpc) if the galaxies approach within 20 kpc of each other. Since M81 is currently at a projected distance
%of 79 kpc, the present-day kinematics of NGC 2976 are probably unaffected by the interaction, but 
%nevertheless, the optical galaxy and the inner HI disk (Stil \& Israel 2002a, 2002b) both appear regular, symmetric, and undisturbed, 
%it has likely interacted with M81 in the past. Appleton et al. (1981) discovered a faint HI streamer stretching from M81 to NGC 2976, implying an interaction between the %two objects. Boyce et al. (2001) used HIJASS data to show that this gas comprises a single tidal bridge that smoothly connects the two galaxies (see their Fig. 2a). The %bridge contains somewhat more HI than NGC 2976 itself ($2.1 \times 10^8 M_{\odot}$ and $1.5 \times 10^8 M_{\odot}$, respectively). 
In summary, after the proto-structure decouples from the expanding background and turns around, the growth of angular momentum is reduced to a second order effect, but subsequent interactions (e.g., the one between M31 and NGC 2976) have an important role in the overall evolution of the total angular momentum of the
protostructure. This is confirmed by the results obtained from interacting systems, where the spin is visibly perturbed by the presence
of a nearby companion.
Tidal interaction of a dwarf with a neighbor in later phase of evolution of the dwarfs 
can increase (but see also Cervantes-Sodi 2010a) the effects of tidal torquing in the phase of proto-dwarf which furnish the angular momentum to the dwarf itself. 
In the present paper, we did not considered the tidal interaction among M31 and NGC 2976, which would further shallow its density profile.

\subsubsection{NGC 5963 and NGC 5949}

In Figs. 5a, b, we studied two other galaxies, NGC 5963 and NGC 5949. 
NGC 5963 is a SC galaxy in the NGC 5866 group (Fouqu\'e et al. 1992). At a projected distance of 430 kpc, there is another galaxy of the group, namely NGC 5907,
distance so large to lead us think that NGC 5963 is not interacting with NGC 5907, and similar considerations lead to conclude that it 
is currently not interacting with its neighbors.
NGC 5963 has a similar blue magnitude as NGC 2976 but has a mass four times larger ($1.4 \times 10^{10} M_{\odot}$). 
From the center outside, the galaxy is characterized by a bar-like structure, outside a disk-like region and a LSB, approximately exponential disk.
%The luminous component of NGC 5963 does not contain an easily identifiable exponential disk. At the center of the galaxy there is a bright, elongated, bar-like feature, and 
%outside there is a small disk-like region (500 pc in radius). At a radius of 15" (950 pc) the surface brightness profile begins a steep
%decline, falling by nearly 3 mag over 18". Surrounding this region is an LSB, nearly exponential disk that extends out to a radius of at
%least 120". 
Usually, it is assumed that the central region of NGC 5963 is a bulge at the center of a very faint disk (e.g., Kormendy \& Kennicutt
2004).
%It is difficult to interpret NGC 5963 surface brightness profile in terms of the standard model of a disk galaxy. Ordinarily, one
%might assume that the bright central region of NGC 5963 is a bulge that just happens to be at the center of an unusually faint
%disk or that this structure is a pseudo-bulge that has formed via secular evolution of the galaxy (e.g., Kormendy \& Kennicutt
%2004). 
Non-circular motions are present in the galaxy (S03).
\begin{figure*}
\centering
\subfigure{\includegraphics[width=15.4cm]{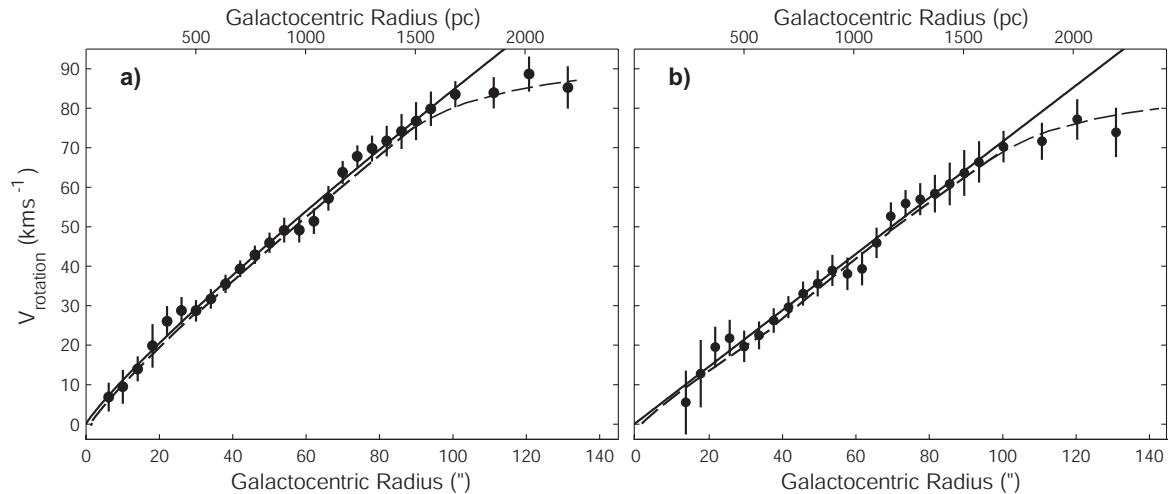}}  \goodgap
\caption{
Panel (a). Minimum disk rotation curve of NGC 2976.  
Black circles with error-bars represent the rotation velocities relative to the minimum disk obtained by S03.
The solid line is a power-law fit to the rotation curve which corresponds to a density profile of 
$\rho \propto r^{-0.27}$, obtained by S03. The dashed line plots the result of the model of the present paper. 
(b) Maximum disk rotation curve of NGC 2976. Similarly to panel (a), but in this case $\rho \propto r^{-0.01}$.
%The dark matter rotation velocities are displayed as black circles with error-bars, as obtained by S03.
%The solid black curve is a power-law fit to the halo velocities
%(for $14" < r < 105"$), which corresponds to a density profile of $\rho \propto r^{-0.01}$. The dashed line is again the result of the model of the present paper.
}
\end{figure*}

NGC 5949 is a SBc galaxy with similar profiles of surface brightness to those of NGC 2976 (S03), even if NGC 5949 has a larger blue magnitude with respect to
NGC 2976, namely $M_B=-18.2$, and also a larger mass ($M \simeq 10^{10} M_{\odot}$). Another similitude between the two objects is that both have a nucleus, a shallow (steep) inner disk (outer disk).

\begin{figure*}
\centering
\subfigure{\includegraphics[height=130mm,width=180mm]{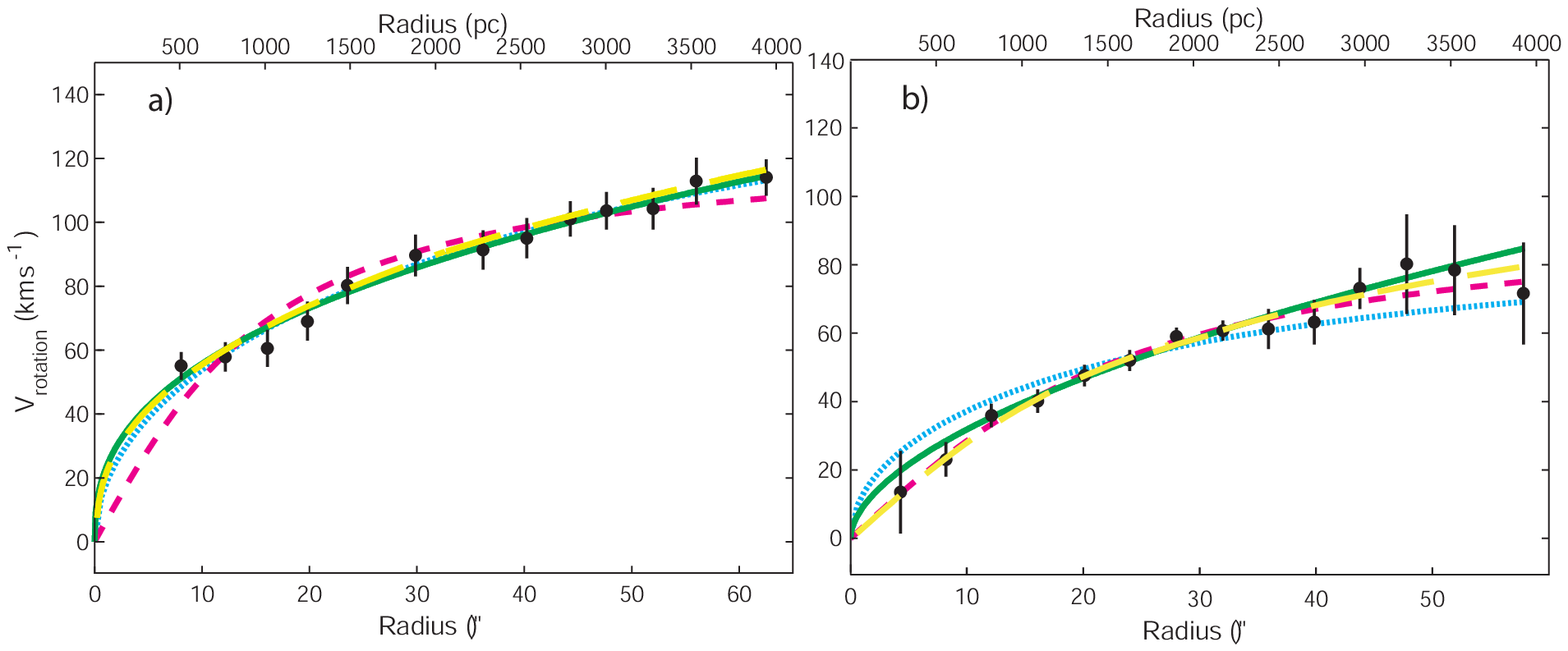}}  \goodgap
\caption{
Panel (a). Disk-subtracted rotation curve of NGC 5963.
% (for $M_{\ast}/L_R=1.24 M_{\odot}/L_{\odot, R}$). 
Similarly to Fig. 4, the black points represent the dark matter halo rotation curve, obtained by S05, after subtracting the stellar disk. The thick green, the dotted cyan, the short-dashed magenta,  show, respectively, the power-law, NFW, and pseudo-isothermal fits obtained by S05, respectively.  
The long-dashed yellow curve is the dark matter halo rotation curve obtained from the model in the present paper. Panel (b). Disk-subtracted rotation curve of NGC 5949. 
%(for $M_{\ast}/L_R=1.64 M_{\odot}/L_{\odot, R}$). 
Symbols are as in panel (a).
}
\end{figure*}
The procedure to obtain rotation curves and dark matter density profiles of the two dwarfs is the same to that of NGC 2976 (S03). 
The limits placed by S05 on $\alpha_{DM}$ are $\alpha_{DM}=0.79 \pm 0.17$, $\alpha_{DM}=0.93 \pm 0.04$ for the maximum and minimum disk of NGC 5949 and
$\alpha_{DM}=0.75 \pm 0.10$, $\alpha_{DM}=1.41 \pm 0.03$ for the maximum and minimum disk of NGC 5963. The dark matter density profile slopes derived after subtracting the stellar disks are $\alpha_{DM}=0.88$ for NGC 5949, and $\alpha_{DM}=1.20$ for NGC 5963, respectively.
In Fig. 5a, the black points are the dark matter halo rotation curve, obtained by S05, after subtracting the stellar disk. The thick green, the dotted cyan, the short-dashed magenta,  show, respectively, the power-law, NFW, and pseudo-isothermal fits obtained by S05, respectively.  
The long-dashed yellow curve is the dark matter halo rotation curve obtained from the model in the present paper. 
%the black points are the dark matter halo rotation curve of NGC 5963 after subtracting the stellar disk ($M_{\ast}/L_R=1.24 M_{\odot}/L_{\odot,R}$).  
%The thick green, the dotted cyan, the short-dashed magenta, and the long-dashed yellow curves show power-law, NFW, pseudo-isothermal, and the curve obtained
%from the model in the present paper, fits to the halo rotation curve, respectively.
NGC 5963 is characterized by a very steep central density profile. S05 showed that a power-law fit with $\alpha_{DM}=1.20$ and a NFW fit with $c=14.9$ and $r_s \simeq 11$, represents very well the rotation curve.  
%quite clearly has a very steep central density profile. A power law with a slope of $\alpha_{DM}=1.20$ fits the rotation curve very well, and an NFW fit with $r_s \simeq %11$ kpc and a concentration parameter of 14.9 is nearly as good.  
A pseudo-isothermal model gives a much worse fit\footnote{Differently from Fig. 1d, in Fig. 5a,b we compared the results to the NFW profile to be consistent with S05, who compared his rotation curves to the NFW profile.}. 
The long-dashed yellow curve obtained with the model of this paper is an excellent fit. In order to obtain the quoted good fit, inside $\simeq 100$ pc we not also had to reduce the magnitude of $h$ but also that of random angular momentum, $j$, (similarly to what done in Williams et al. 2004). 
In fact, as shown in Fig. 1d, one obtain the steepest slope in the case the proto-structure is characterized by $h=0$ and baryon fraction $F_b=0$, as in dissipationless N-body simulations. The density profile is approximated by an Einasto profile. This last  has a slope of $\simeq -0.8$ at $\simeq 100$ pc and $\simeq -1.4$ and at $\simeq 1 $kpc. If the proto-structure contains baryons and the ordered angular momentum is not zero the density profile will be flatter (as seen in Fig. 1, and Fig. 2). 
In the case of NGC 5963, the baryon fraction is obtained as in the case of NGC 2976, by means of Mcgaugh et al. (2010), for an objet having $V_c \simeq 120$ km/s, 
and is given by $f_d \simeq 0.12$.
As always reported, a good fit to NGC 5963, is obtained  
with a value of specific angular momentum $h_{\ast}/2$ and reducing the random angular momentum $j$ to $j/2$ (see Fig. 6 of Williams et al. 2004), which steepens the density profile.  
In Fig. 5b, we plot the disk-subtracted rotation curve of NGC 5949. 
%(for $M_{\ast}/L_R=1.64 M_{\odot}/L_{\odot,R}$). 
Symbols are as in panel Fig. 5a.
A power-law fit with $\alpha_{DM}=0.88$, or a PI density profile, give good fits to the density profile, while a NFW model fit has the problem that fit parameters are not usefully constrained (S05).
%NFW fits to the disk subtracted rotation curve can be carried out, but the fit parameters are not usefully constrained. 
%NGC 5949 is best fit by a PI density profile, but power laws with slopes slightly shallower than NFW ($\alpha_{DM}=0.88$), and the modified pseudo-isothermal profile of %equation:
%\begin{equation}
%\rho=\frac{\rho_c}{(r/r_c)^{1/2}(1+r/r_c)^{3/2}}
%\label{eq:pseud}
%\end{equation}
%where $r_c$ and $\rho_c$ are the characteristic radius and density, respectively,
%fit very well. 
The long-dashed yellow curve obtained with the model of this paper is an excellent fit. In this case, $V_c \simeq 80$ km/s, and $f_d \simeq 0.1$
and the value of specific ordered angular momentum $h_{\ast}/1.6$.
Before going on, we want to recall that S03 and S05 analyses method differs in a few aspects from the other precedent cited studies.
Firstly, their inner slope values are derived from a single power law fit to the entire rotation curve. Secondly, most of their
models do not take into account the gas component due to a lack of HI observations. 
However, the stellar disk has much more mass than gas, at the center of the studied galaxies. As shown by Bolatto et al. (2002), and S03, if one takes into account also the gas, one obtains a change in $M/L$ of $\simeq 20\%$.
%In the galaxies that they studied, however, the stellar disk almost always contributes significantly more mass at the center of the
%galaxy than the gas does, and the effect of including the gas is similar to a 20\% change in the stellar mass-to-light ratio (Bolatto
%et al. 2002; S03). 
%{\it 
The slope of the dark matter density profile decreases of a factor 2\%--12\%, if one arbitrarily increases the stellar mass-to-light ratio ($M_{\ast}/L$) to simulate this effect.
%, the slope of the dark matter density profile decreases by 2\%--12\%. 
As shown by S05 (section 2.2), if one allows that stellar disk have a non-zero thickness, one obtains an offset of this decrease.

Finally, I would like to add that the same kind of processes operating in the late-type dwarfs that we previously discussed take place in LSBs, 
and that LSBs case was already studied in DP09 (Fig. 4).

\subsubsection{Explaining the differences}

The quoted examples show that not all dwarfs are characterized by flat inner cores (like NGC 2976), some have intermediate slopes between pseudo-isothermal profiles and NFW profiles (NGC 5949) other are well fitted by NFW profile (NGC 5963). 
One question we could ask is: why NGC 5949, NGC 5963 (and also NGC 4605; NGC 6689) have steeper inner profiles, in some case (NGC 5963) compatible with the NFW model?
In Sect. 3.1, we have seen how larger values of the baryons fraction and tidal interaction with neighbors produces flatter profiles. 
However, Fig. 1 and Fig. 2 show that an increase in angular momentum has a role more important than an increase in baryon fraction (increase of a factor 2 in angular momentum 
produces similar effects than increase of a factor 3 in baryon fraction).  
For what concerns baryon fraction, the baryon fraction in the case of NGC 5963 is $f_d \sim 0.12$ and it is similar in the case of the other two galaxies (NGC 2976, 
NGC 5949). 
One should take into account that it is not only important if the baryons fraction is similar for different dwarfs (as in the case of NGC 2976, NGC 5949, NGC 5963) to have similar density profiles (all the rest being the same), but is very important how it is distributed in the galaxy (e.g., Hoeft et al. 2006, show in their figure 5 how the baryons fractions can strongly change going from the center to the outskirts of the dwarf). Now, the inner baryon fraction of NGC 2976 and NGC 5963 (using S03, S05 data and Bosma et al. 1988 data), at 2 kpc, are similar, and of the order of $f_b$.  
This very high inner baryon fraction are due to a funneling of baryons toward the center of the dwarfs. 
In fact, in the case of NGC 2976, Williams et al. (2010) results 
%found similar ancient populations at all radii but significantly different young populations at increasing radii. In particular, outside of a well-measured break in the %disk surface brightness profile, the age of the youngest population increases with distance from the galaxy center, 
suggest that star formation is shutting down from the outside-in.   
%The current rate and gas density suggest that rapid star formation in NGC 2976 is currently in the process of ceasing from the outside-in due to gas depletion.
It is possible that depletion of gas from the outer part of the disk, and formation of stars in the central part of the galaxy arose because of an interaction of NGC 2976 with the core of M81 group more than 1 Gyr ago. This last event, could have stripped halo gas and then produced an inflow of it from outer disk to the center of galaxy (Williams et al. 2010). 
%This process of outer disk gas depletion and inner disk star formation was likely triggered by an interaction with the core of the M81 group $\geq$ 1 Gyr ago that stripped %the gas from the galaxy halo and/or triggered gas inflow from the outer disk toward the galaxy center (Williams et al. 2010).
So, in NGC 2976, the funneling of baryons toward the central region produces a larger exchange of angular momentum through baryons and dark matter and an inner flatter profile
with respect to other dwarfs. 
We know that the surface brightness profile of NGC 5949 is very similar to that of NGC 2976, and probably it has similar inner baryon fraction of NGC 2976.  
For what concerns, NGC 5963, differently from the other galaxies, inner and outer surface brightness profiles are characterized by a sharp transition. As a consequence,  
the derived stellar disk rotation curve have a peak at small radii followed by a steep fall, with the consequence that the galaxy has outer parts dominated by dark matter 
(see Fig. 6b of S05).
%For what concerns NGC 5963, differently from the other galaxies it has a sharp transition between the inner and outer surface brightness profiles. This causes the derived %stellar disk rotation curve to peak at small radii ($r =20"$) and then drop steeply, making the outer parts of the galaxy highly dark matter
%dominated (see Fig. 6b of S05). 
As previously discussed, the galaxy could contain a pseudo-bulge, and so like in NGC 2976 some process 
have funneled baryons toward the center of the galaxy.
This means that in the case of these three dwarfs the baryon fraction is playing a similar role in shaping the density profile. 
According to our model, another factor in shaping density profile is the mass of the dwarf.
Among the galaxies studied by S05, namely NGC 2976, NGC 4605, NGC 5949, NGC 5963, and NGC 6689, NGC 5963 has the
highest central surface brightness and the most unusual surface brightness profile.
% (see S05, sect. 3.5.3). 
%NGC 5963 is at the brighter end of the luminosity range of the quoted galaxies ($M_I=-19.1$). 
It does also have the highest mass, in the sample and the highest magnitude ($M_I=-19.1$).
From the point of view of our model, a larger mass imply a steeper profile (however in the case of NGC 5963, NGC 5949, and NGC 2976 this difference is very small (see Table 7 of S05). Also in a recent paper, dB08, by using the THINGS sample, showed that in smaller objects core dominated halo is clearly preferred over a cusp-like halo, while for massive, disk dominated galaxies, PI and NFW fit apparently equally well.
In other terms, for low mass galaxies a core dominated halo is clearly preferred over a cusp-like halo, while for massive, disk dominated galaxies, all halo models give fits of similar quality.
%fit apparently equally well. 
This because more massive galaxies tend to have a more compact light-distribution, and therefore a more compact stellar mass distribution as well. This makes the contribution of the disk to the inner rotation curve critically dependent on the stellar mass-to-light ratio, and the initial mass function (IMF). So one could imagine a situation where the $M/L_{\ast}$ could be changed to produce flat profiles for massive galaxies, and these changes do not have to be very large. 
Low mass galaxies have instead a reduced sensitivity to the $M/L_{\ast}$ values, implying that changes of the quoted ratio does not influence the profile shape. 

For what concerns, the role of tidal torques, we know that in the case of NGC 2976 we need a higher value of the specific angular momentum ($h$, $\lambda=0.04$) than in the case of NGC 5963 and NGC 5949 to fit the profile. Tidal torques influence not only dwarfs during their formation, but if the environment contains close galaxies, tidal torques can influence the shape of the density profile. A search in NED (Nasa Extragalactic Database), concerning NCG5963 and the other quoted galaxies, gives to us the following results: as previously reported, NGC 2976 is a satellite of M81, 
NGC 5949 closest neighbor is a dwarf galaxy 300 kpc away. 
NGC 5963 is part of a small group; the closest neighbor is a dwarf galaxy 200 kpc away, and the massive galaxy NGC 5907 ($3\times 10^{10} M_{\odot}$, Lequeux et al. 1998) is about 400 kpc away. So, NGC 5963 is the only other galaxy that could be regarded as having a similar environment to NGC 2976, even if the actual projected distance between M81 and NGC 2976 is 79 kpc and in the case of NGC 5963 the neighbors are at a much larger distance. Moreover,
%and moreover 
although for NGC 2976 there is good evidence that it has interacted with M81, there is not really significant evidence for an interaction between NGC 5963 and its neighbors.
%\footnote{NGC 5907 has long been used as the prototype of a "noninteracting" warped galaxy. The closest dwarf companion galaxy of NGC 5907 is PGC 54419, which is projected %to be only 36.9 kpc from the center of NGC 5907. This dwarf is seen at the tip of the HI warp and in the direction of the warp. Hence, NGC 5907 is then probably interacting %with PGC 54419 but not with NGC 5963, too far away}.   
In other terms the galaxy can be considered isolated and, as seen in Sect. 3.2, this imply that one cannot expect flattening of the density profile by tidal interactions, like for NGC 2976. Similar conclusions are valid for NGC 5949. Differently, in the case of NGC 2976, the interaction with M81 further contributed to the flattening of the density profile. 

In an attempt to have a deeper insight on the issue (inner profile shape and environment), we have analyzed the rotation curves of a sample of dwarfs, LSB, and three galaxies from THINGS. 
We fitted the rotation curves of the dark halo with power laws, as did by S05 and Oh10 and related each galaxy to the so-called tidal or galactic isolation index, TI, whose negative values correspond to isolated\footnote{Isolated galaxies are galaxies located in low density environments and selected according to fixed criteria (see Karachentseva 1973; Karachentsev \& Petit 1990; Vavilova et al. 2009)}
galaxies of the general field and positive values to members of the groups.
The quoted index was introduced by Karachentsev \& Kashibadze (2006) in their study of the masses of the Local Group and of the M81 group, obtained by distortions in the local velocity field. It is defined as: 
\begin{equation}
\Theta=max[\log (M_k/D^3_k)]+C 
\end{equation}
where ($M_k;D_k$) are the mass and distance to the chosen galaxy of the $k$ - th system and C is a normalization constant.
Estimating $\Theta$ for a given galaxy is actually quite difficult and typically
possible only for systems in the Local Group. A cross match between our sample and the Catalog
of Neighboring Galaxies (Karachentsev et al. 2004) and  Karachentsev and Kashibadze (2006) allows us to infer $\Theta$ for the sample.

The result is plotted in Table 1. The first column identifies the galaxy, the second the type of galaxy (dwarfs (type=0), LSB (type=1) and the three larger galaxies from THINGS (type=2), and the third the paper from which it was taken, last column is the so called tidal index, $\Theta$, defined as (Karachentsev et al. 2004; Karachentsev and Kashibadze 2006).
%\begin{equation}
%\Theta=max[\log (M_k/D^3_k)]+C 
%\end{equation}
%where ($M_k;D_k$) are the mass and distance to the chosen galaxy of the $k$ - th system and C is a normalization constant.

In the case of NGC 2976, we have $TI=2.7$, indicating that the galaxy is non-isolated, and as previously seen its inner slope has $\alpha \simeq 0.01$. In the case of NGC 4605, we have $TI=-1.1$, indicating that the galaxy is isolated, and the inner slope is steeper ($\alpha \simeq 0.78$). 
%Several other similar example are possible, showing flat profiles for non-isolated dwarfs.
In the case of NGC 2366, we have $TI=1$, indicating that the galaxy is non-isolated, and the inner slope is $\alpha \simeq 0.32$ (Oh10). 
%In the case of ICG2574, we have $TI=0.9$, indicating that the galaxy is non-isolated, and $\alpha \simeq 0.09$ (Oh et al. 2010). 
As reported in the introduction, one of the most recent studies on mass modeling for dark matter component of galaxies is the work of Oh10, in which all the seven dwarfs chosen from THINGS are better described by core-like models. These galaxies are included in Table 1. However, we want to point out that at least 5 of the quoted dwarfs are characterized by a positive TI, namely NGC 2366 ($T=1$), ICG2574 ($T=0.9$), HoII  ($T=0.6$), DDO53 ($T=0.7$), HOI ($T=1.5$), indicating that the objects are non-isolated, and one should expect flat profiles, as observed. It would be interesting to extend the previous set of Oh10 to include isolated objects and to compare the density profile of the last with those of non-isolated dwarfs. 

From Table 1, is evident that, in the limit of the small sample considered,
%although with a scatter hard to quantify because of the small sample, 
$\alpha$ correlates with $\Theta$.
%for dwarfs galaxies. 
Moreover, the large value of the Spearman rank correlation coefficient, being $C(\alpha; \Theta) =-0.81$ (for dwarfs case), strongly suggests that the trend observed is not an artifact due to small statistics.
%The LSB sample is too small to say anything, but we note that one of the systems follows quite
We want to add that we are working to build up a bigger sample (also using the WHISP sample by Swaters et al. 2002), and and make fits with power laws, and generalized NFW models. 
%In order to enlarge the sample we will resort to two other commonly statistics used to quantify whether a galaxy lives in a dense environment or is rather an isolated one %(to be discussed in the forthcoming paper). 
In order to further test the possible correlation between the inner slope and the environment, we will resort to two other commonly used indicators. The first, denoted as $\Sigma_{10}$, is the projected number density of galaxies inside the circle of radius equal to the projected distance of the tenth nearest neighbor (Dressler et al. 1980), while the second, denoted as $D_{750}$, is the projected number density of galaxies within a circle of radius equal to 750 kpc centred on the system of interest (Treu et el. 2009). $\Sigma_{10}$ and $D_{750}$ positively correlate with $\Theta$ so that we can use them as supplementary tools to check whether the environment impacts the inner slope of the density profile.

\begin{table}
%\tiny
%\scriptsize
\caption{Dwarf (type = 0), LSB (type = 1), and normal spirals (type=2) sample and constraints on model parameters.}
\begin{center}
\begin{tabular}{ccccccccc}
\hline
Id & Type & Ref & $\Theta$  & $\alpha$  \\
\hline
UGC1281	& 1 & dBB02 & -1.2 &  $0.92_{-0.08}^{+0.08}$   \\
UGC8490 & 0 & Sw11 & -1.1 & $0.89_{-0.06}^{+0.06}$   \\
NGC4605	& 0 & S05 &	-1.1 & $0.78_{-0.06}^{+0.06}$   \\
M81dwB & 0 & Oh10 & -0.9 &  $0.39_{-0.06}^{+0.06}$  \\	
NGC5023	& 1 & dBB02 & -0.5 & $0.79_{-0.07}^{+0.07}$  \\
NGC4736 & 2 & dBB02 & -0.5 & $0.81_{-0.08}^{+0.08}$ \\
NGC3274	& 1 & dBB02 & -0.3 & $0.75_{-0.07}^{+0.07}$  \\
NGC2403 & 2 & dBB02 &  0 & $0.51_{-0.06}^{+0.06}$ \\
UGC7524	& 0	& Sw11 & 0.1 &  $0.46_{-0.06}^{+0.06}$  \\
UGC7559	& 0	& Sw11 & 0.1 & $0.38_{-0.06}^{+0.06}$   \\
NGC7793 & 2 & dBB02 &  0.1 & $0.07_{-0.05}^{+0.05}$ \\
HoII & 0 & Oh10 & 0.6 & $0.43_{-0.06}^{+0.06}$   \\
DDO53 & 0 &	Oh10 & 0.7	& $0.38_{-0.06}^{+0.06}$   \\
IC2574 & 0 & dBB02 & 0.9 &  $0.09_{-0.07}^{+0.07}$   \\
NGC2366 & 0 & dBB02 & 1.0 & $0.32_{-0.1}^{+0.1}$  \\
HoI & 0 & Oh10 & 1.5 &  $0.38_{-0.06}^{+0.06}$  \\
NGC2976	& 0	& dBB02 & 2.7 &  $0.01_{-0.1}^{+0.1}$  \\
\\
Sw11 & (Swaters et al. 2011) \\

\hline
\end{tabular}
\end{center}
\end{table}

The quoted example and the discussion of the previous sections, indicates that tidal interaction, baryon fraction, and environmental effects, influence the slope of the dwarfs density profile. This does not come unexpected, since we know that on the other hand, the morphology--density relation illustrates that environment is an important factor. 
The existence of a correlation between galaxy morphology and local density environment (Dressler 1980, Goto et al. 2003, Park et al. 2007) is a motivation to search for  correlations between local environment and internal properties of galaxies. One of this correlation of interest to us, is the possible correlation between $\lambda$ and 
environment.
Concerning this issue, there are different opposite results. Macci\'o et al. (2007) found an absence of correlation between $\lambda$ and environment in cosmological simulations, although the opposite was found 
by other studies using similar techniques (e.g. Avila-Reese et al. 2005, Cervantes-Sodi et al. 2008a,b, 2010a, 2010b). 
Avila-Reese et al. (2005) employed a $\Lambda$CDM N-body simulation to study the properties of galaxy-size dark matter
haloes as a function of global environment, where $\lambda$ was one of the studied properties of the halo.
%, in particular its value in different environments such as clusters, voids or field.
Cervantes-Sodi found a marked correlation of $\lambda$ with mass (Cervantes-sodi et al. 2008a, b), a correlation between $\lambda$ and cluster environment (Cervantes-Sodi et al. 2010b), and showed that events such as interaction with close neighbors play an important role in the value of the spin for the final configuration (Cervantes-Sodi et al. 2010b). Even the spin alignment of dark matter haloes in different environments, has been studied in some N-body simulations
(Arag\'on-Calvo et al. 2007). From the observational point of view, there has not been done much study on the effects of environment on the slopes of the dark matter halos,
simply because there are not many large samples needed for that, yet.
%larger samples needed for that do not exist, yet. 
As previously reported, there is a first sample (WHISP by Swaters et al. 2002)
and to this will be added
%hopefully this will change 
in the next few years, the LITTLE THINGS (PI: D. Hunter) and VLA-ANGST (PI: J. Ott), and 
they should give access to high-resolution HI observations from $\simeq 50$ to 60 dwarf galaxies in various environments. 
Moreover, THINGS in that respect is a sample that was specifically selected not
to contain too many interacting galaxies (apart from one or two show case galaxies), and
is therefore not an ideal sample to look at the effects mentioned. Moreover, the techniques
actually used to infer the rotation curve makes difficult to address the problem of the quoted
correlation. The problem is that when a galaxy is undergoing an interaction, the orbits of
the gas particles will change, especially those in the outskirts. So the entire assumption of
the gas going around in circular orbits, and the analysis using the ROTCUR approach, will
break down. One could, of course, still try to analyze the velocity field and, making some
reasonable assumptions, come up with a rotation curve. But if one then find that the inner
slope is smaller, there is the problem of how confident can one be in that result: the low
value of the slope could be due to the fact that we are fitting a model of circular velocities
to what in reality are non-circular motions. One partial solution to the problem would be
that of studying dwarfs non strongly interacting with neighbors and to compare with dwarfs
in voids or to wait for more sophisticated techniques. 
Clearly, many work is needed from the theoretical and observational point of view in order to have deeper insights on the role of environment and density profiles of dwarfs
and other important questions require further research, including the mechanisms governing dwarf galaxy evolution, and the origin of the morphology--density
relation.

\section{Conclusions}

Using the SIM model introduced in DP09, we studied how the shape of density profiles of dwarfs in the mass range $10^8-10^{10} M_{\odot}$ 
changes when angular momentum originated by tidal torques, and baryon fraction are changed. 
As a first step, we calculated the reference density profiles following DP09, plotted in Fig. 1a, then we calculated 
density profiles of haloes having ordered angular momentum in a range of values compatible with dwarf galaxies haloes and baryon fraction as 
obtained by McGaugh et al. (2010). We found that density profiles steepen with increasing mass of the halo while increasing angular momentum the density profiles flatten, in agreement with previous works (Sikivie et al. 1997; Avila-Reese et al. 1998; Nusser 2001; Hiotelis 2002; Le Delliou \& Henriksen 2003; Ascasibar et al. 2004; Williams et al. 2004). In the case tidal torquing is shut down (zero angular momentum) and baryons are not present, the density profile is very well approximated by an Einasto profile.
Increasing baryon fraction one gets flatter profiles. This is due to the fact that when more baryons are present the energy and angular momentum transfer from baryons to DM is larger and DM moves on larger orbits reducing the inner density. We then repeated the same calculations for three dwarfs studied by S03, S05, namely
NGC 2976, NGC 5949, and NGC 5963. NGC 2976 has a core profile, NGC 5963 a cuspy profile and NGC 5949 an intermediate one. We calculated the baryon fraction 
%from the baryon mass and total mass of the dwarfs 
using some assumptions and McGaugh et al. (2010) results and then fitted the rotation curves obtained by S03, S05 with the theoretical curves obtained from our model. NGC 2976 is characterized by a larger value of angular momentum ($\lambda \simeq 0.04$) with respect the other two dwarfs ($\lambda \simeq 0.025$, for NGC 5949; $\lambda \simeq 0.02$, and random angular momentum, $j$, 1/2 of those of the reference haloes (Fig. 1a), for NGC 5963). 
Moreover, there are evidences of past tidal interaction between NGC 2976 and M31. Appleton et al. (1981) discovered a faint HI streamer stretching from M81 to NGC 2976, 
implying an interaction between the two objects, 
%Boyce et al. (2001) used HIJASS data to show that this gas comprises a single tidal bridge that smoothly connects the two galaxies (see their Fig. 2a).
and observations of Williams et al. (2010) showed a  process of outer disk gas depletion and inner disk star formation likely triggered by an interaction with the core of the M81 group more than 1 Gyr ago. This interaction could be another reason why the density profile of NGC 2976 is so flat. 
In the case of NGC 5949 and NGC 5963 there is no evidence of this kind of tidal interaction and the closest neighbors are at several hundreds kpc.
The small sample that we used, shows a correlation between the $\alpha$ and the tidal index $\Theta$.

Concluding, dwarfs galaxies which acquired larger angular momentum by tidal interaction in their formation phase and/or interacted tidally in later phases with neighbors, show a flatter density profile than those who had less tidal interactions.

\section*{Acknowledgments}
%\acknowledgements

We would like to thank Josh Simon, Erwin de Blok, Elias Brinks, Benoit Famaey, Alister Graham, Igor Karachentsev, Valentina Karachentseva, and Bernardo Cervantes-Sodi for stimulating discussions on the topics related to the subject of this paper.

{}

\begin{thebibliography}{}
\bibitem{} Antonuccio-Delogu, V., \& Colafrancesco, S. 1994, ApJ, 427, 72 (ADC)
\bibitem{} Appleton, P. N., Davies, R. D., \& Stephenson, R. J. 1981, MNRAS, 195, 327
\bibitem{} Arag\'on-Calvo M. A., van de Weygaert R., Jones B. J. T., van der Hulst J. M., 2007, ApJ, 655, L5
\bibitem{} Ascasibar, Y., Yepes, G., \& Gottlober, S. 2004, MNRAS, 352, 1109A
%\bibitem{} Athanassoula, E.,  Bosma, A., and Papaioannou, S., 1987, A\&A 179, no. 1-2, pp. 23-40
\bibitem{} Avila-Reese, V., Firmani, C., \& Hernandez, X. 1998, ApJ, 505, 37
\bibitem{} Avila-Reese V., Colin P., Gottlùober S., Firmani C., Maulbetsch C., 2005, ApJ, 634,51
\bibitem{} Babul, A.,  and H. C. Ferguson, H. C., 1996, ApJ 458, no. 1, 100-119.
%\bibitem{} Begeman,  K. G., Broeils, A. H.,  and Sanders,  R. H., 1991, MNRAS 249, p. 523
\bibitem{} Benjamin F. W, et al., 2010, ApJ 709, 135
\bibitem{} Blais-Ouellette, Amram, P., Carignan, C., and Swaters, R., 2004, A\&A 420, no. 1, pp. 147-161
\bibitem{} Blumenthal, G. R., Faber, S. M., Flores, R., \& Primack, J. R. 1986, ApJ, 301, 27
\bibitem{} Boissier, S., Monnier, R. D., van Driel, W., Balkowski, C., \& Prantzos, N. 2003, Ap\&SS, 284, 913
\bibitem{} Bolatto, A. D., Simon, J. D., Leroy, A., \& Blitz, L. 2002, ApJ, 565, 238
\bibitem{} Borriello, A., \& Salucci, P. 2001, MNRAS, 323, 285
\bibitem{} Bothun, D., 1997, AJ 114, no. 5, pp. 1858-1882
\bibitem{} Boyce, P. J., et al. 2001, ApJ, 560, L127
%\bibitem{} Broeils, A. H., 1992, Dark and Visible Matter in Spiral Galaxies, Ph.D. thesis, University of Groningen, Groningen, The Netherlands
\bibitem{} Broeils, A. H. thesis, Univ. Groningen, 1990.
\bibitem{} Bullock, J. S.,  et al. 2001, ApJ 555, 240 
\bibitem{} Burkert, A. 1995, ApJ, 447, L25
\bibitem{} Carignan, C. \& Beaulieu, S., 1989, ApJ. 347, 760-770. 
\bibitem{} Catelan, P., \& Theuns, T. 1996, MNRAS, 282, 436
\bibitem{} Cervantes-Sodi, B., Hernandez, X., Park, C., Kim, J., 2008a, MNRAS 388, 863
\bibitem{} Cervantes-Sodi, B., Hernandez, X., Park, C., Kim, J., 2008b, RevMexAA (Serie de Conferencias) 34, 87-90
\bibitem{} Cervantes-Sodi, B., Hernandez, X., Park, C., 2010a, MNRAS 402, Issue 3, pages 1807-1815
\bibitem{} Cervantes-Sodi, B., Park, C., Hernandez, X., Hwang, H. S., 2010b, arXiv:1008.2832 
\bibitem{} de Blok W. J. G.,  and McGaugh, S. S., 1997, MNRAS 290, no. 3, pp. 533-552
\bibitem{} de Blok, W. J. G. 2003, in ASP Conf. Ser. 220, Dark Matter in Galaxies, ed. S. Ryder, et al. (San Francisco, CA: ASP), 69
%\bibitem{} de Blok, W. J. G.,  and Bosma, A., 2002, A\&A Astronomy 385, no. 3, pp. 816-846
\bibitem{} de Blok, W. J. G., \& Bosma, A. 2002, A\&A, 385, 816
\bibitem{} de Blok, W. J. G., 2010, Advances in Astronomy Volume 2010, Article ID 789293, 14 pages
%\bibitem{} de Blok, W. J. G., and McGaugh, S. S., 1997, MNRAS 290, no. 3, pp. 533-552
%\bibitem{} de Blok, W. J. G., Bosma, A. McGaugh, S., 2003, MNRAS 340, no. 2, pp. 657-678
\bibitem{} de Blok, W. J. G., Bosma, A., \& McGaugh, S. 2003, MNRAS, 340, 657
%\bibitem{} de Blok, W. J. G., Bosma, a., and McGaugh, S., 2003, MNRAS 340, no. 2, pp. 657-678
\bibitem{} de Blok, W. J. G., McGaugh, S. S., Bosma, A., \& Rubin, V. C. 2001, ApJ, 552, L23
\bibitem{} de Blok, W. J. G., Walter, F., Brinks, E., Trachternach, C., Oh, S-H., and Kennicutt, R. C., AJ  136, 2648
%\bibitem{} de Blok, W.J.G., McGaugh, S.S. 1997, MNRAS, 290, 533
\bibitem{} de Naray, R. K., McGaugh, S. S., and de Blok, W. J. G., 2008, ApJ 676, no. 2, pp. 920-943
\bibitem{} de Naray, R. K., McGaugh, S. S., and Mihos, J. C., 2009, ApJ 692, pp. 1321-1332
%\bibitem{} Dekel, A., \& Woo, J. 2002, preprint (astro-ph/0210454)
\bibitem{} Dekel, A., \& Woo, J. 2003, MNRAS, 344, 1131
\bibitem{} Del Popolo A., 2009, ApJ 698, 2093-2113
\bibitem{} Del Popolo, A., \& Gambera, M. 1996, A\&A, 308, 373
\bibitem{} Dressler A., 1980, ApJ, 236, 351
\bibitem{} Duc, P. A., \& Mirabel, I. F. 1998, A\&A, 333, 813
\bibitem{} Eisenstein, D. J., \& Loeb, A. 1995, ApJ, 439, 250
\bibitem{} El-Zant, A., Shlosman, I., \& Hoffman, Y. 2001, ApJ, 560, 636 
\bibitem{} Ferguson, H. C., Binggeli, B., Astronomy and Astrophysics Review, vol. 6 no. 1-2, 67-122
\bibitem{} Fillmore, J. A., \& Goldreich, P. 1984, ApJ, 281, 1
\bibitem{} Flores R., Primack J.R. 1994, ApJ, 427, L1
\bibitem{} Fouqu\'e, P., Gourgoulhon, E., Chamaraux, P., \& Paturel, G. 1992, A\&AS, 93, 211
\bibitem{} Gao, L., \& White, S. D. M. 2007, MNRAS, 377, 5
\bibitem{} Geha M., Blanton2, M. R., Masjedi M. and West, A. A, 2006, ApJ 653, 240
\bibitem{} Gelato, S., and Sommer-Larsen, J., 1999, MNRAS 303, no. 2, pp. 321-328
\bibitem{} Gentile, G., Salucci, P., Klein, U., \& Granato, G. L. 2007a, MNRAS, 375, 199
\bibitem{} Gentile, G., Salucci, P., Klein, U., Vergani, D., \& Kalberla, P. 2004, MNRAS, 351, 903
\bibitem{} Gentile, G., Tonini, C., \& Salucci, P. 2007b, MNRAS, 378, 41
\bibitem{} Gnedin, N. Y.,  and J. P. Ostriker, J. P., 1997, ApJ 486, no. 2, 581-598
\bibitem{} Gnedin, O. Y., Kravtsov, A. V., Klypin, A. A., \& Nagai, D. 2004, ApJ, 616, 16
\bibitem{} Goto T., Yamauchi C., Fujita Y., Okamura S., Sekiguchi
\bibitem{} Governato, F., et al., 2007, MNRAS 374, Issue 4, pp. 1479-1494
\bibitem{} Governato, F., et al. 2010, Nature 463, 203-206
\bibitem{} Grebel, E. K.,  2001, in ``Dwarf Galaxies and their Environment'', 40th Graduiertenkolleg-Workshop, eds. K.S. de Boer, R.-J. Dettmar, \& U. Klein (Bonn:Shaker Verlag), 45-52
\bibitem{} Gunn, J. E. 1977, ApJ, 218, 592
\bibitem{} Gunn, J. E., \& Gott, J. R. 1972, ApJ, 176, 1
\bibitem{} Hayashi, E., Navarro, J. F., Taylor, J. E., Stadel, J., \& Quinn, T. 2003, ApJ, 584, 541
\bibitem{} Hayashi, E., Navarro, J.F., Power, C., et al., 2004, MNRAS 355, no. 3, pp. 794-812
\bibitem{} Hiotelis, N. 2002, A\&A, 383, 84
\bibitem{} Hoeft, M., Yepes, G., Gottl\"ober, S., Springel, V., 2006, MNRAS 371, no. 1, 401-414 
\bibitem{} Hoeft, M. and Gottl\"ober, S., , Hindawi Publishing Corporation, Advances in Astronomy Volume 2010, Article ID 693968, 16 pages
\bibitem{} Hoffman, Y., \& Shaham, J. 1985, ApJ, 297, 16 (HS)
\bibitem{} Hoyle, F. 1949, in IAU and International Union of Theorethical and Applied Mechanics Symposium, Problems of Cosmological Aerodynamics, ed. J. M.
\bibitem{} Burger \& H. C. van der Hulst (Ohio: IAU), 195
\bibitem{} Jaffe, A. H., et al. 2001, Phys. Rev. Lett. 86, 3475
\bibitem{} Jobin, M. \& Carignan, C., 1990, AJ 100, 648-662. 
\bibitem{} Karachentsev, I. D.,  and O. G. Kashibadze, O. G., 2006, Astrophysics 49, No. 1
\bibitem{} Karachentsev, I. D., et al. 2002, A\&A, 383, 125
\bibitem{} Karachentsev, I.D., Karachentseva, V.E., Huchtmeier, W.K., Makarov, D.I., 2004, AJ, 127, 203
\bibitem{} Kaufmann et al. 2007, MNRAS 375, Issue 1, pages 53-67
\bibitem{} Kazantzidis, S., Abadi, M. G., and Navarro, J. F., 2010, ApJL 720, L62-L66
%\bibitem{} Kent, S.M., 1987, AJ 93, pp. 816-832
\bibitem{} Kleyna, J. T., Wilkinson, M. I., Gilmore, G., \& Evans, N. W. 2003, ApJ, 588, L21
\bibitem{} Klypin, A., Gottl\"ober, S., Kravtsov, A. V., \& Khokhlov, A.M. 1999,ApJ, 516, 530
\bibitem{} Klypin, A., Zhao, H., \& Somerville, R. S. 2002, ApJ, 573, 597
\bibitem{} Knebe, A., and Power, C., 2008, ApJ 678, 621-626
\bibitem{} Kormendy, J., \& Kennicutt, R. C. 2004, ARA\&A 42, 603-683
%in press (preprint: astro-ph/0407343)
\bibitem{} Kormendy, J., and Freeman, K. C., 2004, "Scaling laws for dark matter halos in late-type and dwarf spheroidal galaxies," in Dark Matter in Galaxies, International Astronomical Union Symposium no. 220, pp. 377-397
\bibitem{} Kravtsov, A. V., Klypin, A. A., Bullock, J. S., \& Primack, J. R. 1998, ApJ, 502, 48
\bibitem{} Kravtsov, A.V., Klypin, A. A., Bullock, J. S., and Primack, J. R., ApJ 502, no. 1, pp. 48-58, 1998.
\bibitem{} Le Delliou, M., \& Henriksen, R. N. 2003, A\&A, 408, 27
\bibitem{} Lemson G., Kauffmann G., 1999, MNRAS, 302, 111
\bibitem{} Lequeux, J., Combes, F., Dantel-Fort, M., Cuillandre, J. C., Fort, B., Mellier, Y. 1998  A\&AL 334, L9-L12 
\bibitem{} Lilly, S. J., Tresse, L., Hammer, F., Crampton, D., and Le F\'evre, O., 1995, ApJ 455, no. 1, pp. 108-124
\bibitem{} Lokas, E. W., Kazantzidis, S., Mayer, L., and Callegari, S., arXiv:1011.3357v1 [astro-ph.CO]
\bibitem{} Magorrian, J. 2003, in ESO Astrophysics Symposia, The Mass of Galaxies at Low and High Redshift: Proc. ESO Workshop Venice, ed. R. Bender, A. Renzini (Singapore: Springer), 18
\bibitem{} Flores, R. A., \& Primack, J. R. 1994, ApJ, 427, L1
\bibitem{} Lake, G., Schommer, R. A. \& van Gorkom, 1990, J. H. AJ 99, 547-560. 
\bibitem{} Maccio A. V., Dutton A. A., van den Bosch F. C., Morre B., Potter D., Stadel J., 2007, MNRAS, 378, 55
\bibitem{} Maller, A.H. and Dekel, A.: 2002, MNRAS 335, 487.
\bibitem{} Marchesini,D., D'Onghia, E.,Chincarini, G., Firmani,C., Conconi, P., Molinari, E., \& Zacchei, A. 2002, ApJ, 575, 801
%\bibitem{} Maschenko 2007
\bibitem{} Mashchenko, S., \& Sills, A. 2005, ApJ, 619, 258
\bibitem{} Mashchenko, S., Couchman, H. M. P., \& Wadsley, J. 2006, Nature, 442, 539
\bibitem{} McGaugh, S. S., \& de Blok W. J. G., 1998, ApJ 508, 132
\bibitem{} McGaugh, S. S., Schombert, J. M., de Blok, W. J. G. and Zagursky, M. J., 2010, 708, L14-L17, 
\bibitem{} Moore, B. 1994, Nature, 370, 629
%\bibitem{} Moore, B. 1994, Nature, 370, 629
\bibitem{} Moore, B., Ghigna, S., Governato, F., Lake, G., Quinn, T., Stadel, J., \& Tozzi, P. 1999, ApJ, 524, L19
\bibitem{} Mun$\tilde{o}$z-Mateos et al. 2011, arXiv:1102.1724v1 
\bibitem{} Navarro, J. F., Eke, V. R., and Frenk, C. S., 1996, MNRAS 283, no. 3, pp. L72--L78
\bibitem{} Navarro, J. F., et al. 2004, MNRAS, 349, 1039
\bibitem{} Navarro, J. F., et al. 2010, MNRAS 402, 21
%2008, arXiv:astro-ph/0810.1522
\bibitem{} Nusser, A. 2001, MNRAS, 325, 1397
\bibitem{} Oh, Se-Heon et al. 2010, arXiv:1011.2777 
\bibitem{} Ostriker, J.P., Steinhardt, P. 2003, Science, 300, 1909
\bibitem{} Park C., Choi Y., Vogeley M. S., Gott J. R., Blanton M. R., 2007, ApJ, 658, 898
\bibitem{} Peacock, J. A., \& Heavens, A. F. 1990, MNRAS, 243, 133
\bibitem{} Peebles, P. J. E. 1969, ApJ, 155, 393
\bibitem{} Peebles, P. J. E. 1980, The Large Scale Structure of the Universe (Princeton, NJ: Princeton Univ. Press)
\bibitem{} Percival. W. J., et al. 2001, MNRAS 327, 1297
\bibitem{} Pickering, T. E.,  Impey,  C.D.,  van Gorkom, J. H., and Bothun, G. D.,  1997, AJ 114, no. 5, pp. 1858-1882
\bibitem{} Ricotti, M., 2009, MNRAS 392, L45-L49
\bibitem{} Romano-Diaz, E., Shlosman, I., Hoffman, Y., \& Heller, C. 2008, ApJ, 685, 105
\bibitem{} Ryden, B. S. 1988a, ApJ, 329, 589, (R88)
\bibitem{} Ryden, B. S. 1988b, ApJ, 333, 78
\bibitem{} Ryden, B. S., \& Gunn, J. E. 1987, ApJ, 318, 15 (RG87)
\bibitem{} Sales, L. V., Helmi, A., and Battaglia, G., 2010, Advances in Astronomy, Article ID 194345, 14 pages
\bibitem{} Salucci, P., \& Burkert, A. 2000, ApJ, 537, L9
\bibitem{} Schaefer, B. M., 2009, Int. J. Modern Phys. D, 18, 173
\bibitem{} Sellwood, J. A. \& McGaugh, Stacy S. 2005, ApJ, 634, 70
\bibitem{} Sikivie, P., Tkachev, I. I, \& Wang, Y. 1997, Phys. Rev. D, 56, 1863
\bibitem{} Simon, J. D., Bolatto, A. D., Leroy, A., \& Blitz, L. 2003, ApJ, 596, 957
\bibitem{} Simon, J. D., Bolatto, A. D., Leroy, A., Blitz, L., and Gates, E. L., 2005 ApJ 621 757
\bibitem{} Simon, J. D., Bolatto, A. D., Leroy, A., \& Blitz, L. 2004, in ASP Conf. Ser. 32, Satellites and Tidal Streams, ed. F. Prada, D. Martinez-Delgado, \& T. J. Mahoney (San Francisco, CA: ASP), 18
%\bibitem{} Sofue,  Y., and Rubin, V., 2001, Annual Review of Astronomy and Astrophysics 39, no. 1, pp. 137-174, 2001.
\bibitem{} Spano, M., Marcelin, M., Amram, P., Carignan, C., Epinat, B., and Hernandez, O., 2008, MNRAS 383, no. 1, pp. 297-316
\bibitem{} Spekkens, K., Giovanelli, R., \& Haynes, M. P. 2005, AJ, 129, 2119
\bibitem{} Spergel, D. N., Verde, L., Peiris, H. V., et al. 2003, ApJS, 148, 175
\bibitem{} Sprayberry, D., Impey, C. D., Bothun, G. D., and Irwin, M. J., 1995, AJ 109, no. 2, pp. 558-571
\bibitem{} Stadel, J., Potter, D., Moore, B., Diemand, J., Madau, P., Zemp, M., Kuhlen, M., \& Quilis, V. 2009, MNRAS 398, 21
%arXiv:astro-ph/0808.2981
%\bibitem{} Stil, J. M., \& Israel, F. P.  2002b, A\&A, 389, 42
%\bibitem{} Stil, J. M., \& Israel, F. P. 2002a, A\&A, 389, 29
\bibitem{} Stil, J. M., and Israel, F. P., 2002, A\&A 389, 29-41 
\bibitem{} Stoehr, F., White, S. D. M., Tormen, G., \& Springel, V. 2002, MNRAS, 335, L84
\bibitem{} Swaters R. A., van Albada, T. S., van der Hulst, J. M. and R. Sancisi, R., A\&A 2002, 390, 829 
\bibitem{} Swaters, R. A., Verheijen, M. A. W., Bershady, M. A., and Andersen, D. R., 2003, ApJ 587, no. 1, pp. L19-L22
\bibitem{} Swaters, R.A., Madore, B.F., van den Bosch, F. C., and Balcells, M., 2003, ApJ 583, no. 2, pp. 732-751
\bibitem{} Swaters, R.A., Sancisi, R., van Albada, T.S., van der Hulst, J.M.,  ApJ, 729, 118, 2011 (Sw11)
\bibitem{} Treu, T., Gavazzi, R., Gorecki, A., Marshall, P.J., Koopmans, L.V.E. et al., 2009, ApJ 690, 670
\bibitem{} Turner, M. S., 2002, ApJ 576, L101--L104
%\bibitem{} van Albada, T. S., and Sancisi, R., 1986, Philosophical Transactions of the Royal Society A 320, no. 1556, pp. 447-464
\bibitem{} van den Bosch F. C., Burkert A., Swaters R. A., 2001, MNRAS 326, 1205
\bibitem{} van den Bosch, F.C., Robertson, B.E., Dalcanton, J. J., and de Blok, W.J.G., 2000, AJ 119, no. 4, pp. 1579-1591
\bibitem{} Vitvitska, M., Klypin, A. A., Kravtsov, A. V., Wechsler, R. H., Primack, J. R. and Bullock, J. S., 2002 ApJ 581 799
\bibitem{} White, S. D. M. 1984, ApJ, 286, 38
\bibitem{} White, S. D. M., \& Zaritsky, D. 1992, ApJ, 394, 1
\bibitem{} Williams, B. F., et al. 2010, ApJ 709, 135-148
\bibitem{} Williams, L. L. R., Babul, A., \& Dalcanton, J. J. 2004, ApJ, 604, 18
\bibitem{} Yun, M. S., Ho, P. T. P., \& Lo, K. Y. 1994, Nature, 372, 530
\bibitem{} Zaroubi, S., \& Hoffman, Y. 1993, ApJ, 416, 410
\end{thebibliography}
\end{document}